\documentclass{article} 
\usepackage[table,dvipsnames]{xcolor}
\usepackage{iclr2025_conference,times}

\usepackage{amsmath}
\usepackage{amsfonts}
\usepackage{bm}

\def\eqref#1{equation~\ref{#1}}

\def\1{\bm{1}}

\def\vone{{\bm{1}}}

\def\vc{{\bm{c}}}

\def\ve{{\bm{e}}}

\def\vp{{\bm{p}}}

\def\vu{{\bm{u}}}

\def\vx{{\bm{x}}}

\def\vz{{\bm{z}}}

\def\mI{{\bm{I}}}

\def\mQ{{\bm{Q}}}

\DeclareMathAlphabet{\mathsfit}{\encodingdefault}{\sfdefault}{m}{sl}
\SetMathAlphabet{\mathsfit}{bold}{\encodingdefault}{\sfdefault}{bx}{n}

\def\gS{{\mathcal{S}}}

\newcommand{\E}{\mathbb{E}}

\newcommand{\R}{\mathbb{R}}

\usepackage{mathtools}
\usepackage{hyperref}
\hypersetup{colorlinks=true,allcolors=[rgb]{0.2,0.2,0.75}}
\usepackage{url}
\usepackage{microtype}
\usepackage{subfigure}
\usepackage{wrapfig}
\usepackage{graphicx}
\usepackage{bbding}
\usepackage{amssymb}
\usepackage{amsmath}
\usepackage{amsfonts}
\usepackage{pifont}
\usepackage{booktabs}
\usepackage{multicol}
\usepackage{multirow}

\usepackage{listings}
\definecolor{codegreen}{rgb}{0,0.6,0}
\definecolor{codegray}{rgb}{0.5,0.5,0.5}
\definecolor{codeRoyalBlue}{rgb}{0.58,0,0.82}
\definecolor{backcolour}{rgb}{0.95,0.95,0.92}
\lstdefinestyle{mystyle}{
    backgroundcolor=\color{backcolour},   
    commentstyle=\color{codegreen},
    keywordstyle=\color{magenta},
    numberstyle=\tiny\color{codegray},
    stringstyle=\color{codeRoyalBlue},
    basicstyle=\ttfamily\scriptsize,
    breakatwhitespace=false,         
    breaklines=true,                 
    captionpos=b,                    
    keepspaces=true,                 
    numbers=left,                    
    numbersep=5pt,                  
    showspaces=false,                
    showstringspaces=false,
    showtabs=false,                  
    tabsize=2
}
\newcommand{\cat}{\text{Cat}}
\newcommand{\mask}{\text{[MASK]}}

\vspace{-0.3cm}
\title{Reaction-conditioned De Novo Enzyme Design with GENzyme}
\vspace{-0.5cm}

\author{%
  Chenqing Hua$^{1,3,\dag}$\thanks{Corresponding author. $\dag$ Co-authorship.} \ \ \ \ \ \ Jiarui Lu$^{3,4,\dag}$ \ \ \ \ \ \ Yong Liu$^{5}$ \ \ \ \ \ \ Odin Zhang$^6$  \ \ \ \ \ \ \vspace{-0.4cm} \And Jian Tang$^{3,4}$ \ \ Rex Ying$^7$ \ \ Wengong Jin$^{8,9}$ \ \ Guy Wolf$^{3,4}$ \ \ Doina Precup$^{1,3,10*}$ \ \ Shuangjia Zheng$^{2*}$
    \\
    $^1$McGill; \ \ $^2$SJTU; \ \ $^3$Mila-Quebec AI Institute; \ \ $^4$UdeM; \ \ $^5$HKUST; \\$^6$Institute for Protein Design, UW; \ \ $^7$Yale; \ \ $^8$Northeastern; \ \ $^9$Board \& MIT; \ \ $^{10}$DeepMind  \\
}

\iclrfinalcopy 
\begin{document}

\maketitle

\vspace{-0.3cm}
\begin{abstract}
\vspace{-0.2cm}
The introduction of models like RFDiffusionAA, AlphaFold3, AlphaProteo, and Chai1 has revolutionized protein structure modeling and interaction prediction, primarily from a binding perspective, focusing on creating ideal lock-and-key models. However, these methods can fall short for enzyme-substrate interactions, where perfect binding models are rare, and induced fit states are more common. To address this, we shift to a functional perspective for enzyme design, where the enzyme function is defined by the reaction it catalyzes. Here, we introduce \textsc{GENzyme}, a \textit{de novo} enzyme design model that takes a catalytic reaction as input and generates the catalytic pocket, full enzyme structure, and enzyme-substrate binding complex. \textsc{GENzyme} is an end-to-end, three-staged model that integrates (1) a catalytic pocket generation and sequence co-design module, (2) a pocket inpainting and enzyme inverse folding module, and (3) a binding and screening module to optimize and predict enzyme-substrate complexes. The entire design process is driven by the catalytic reaction being targeted. This reaction-first approach allows for more accurate and biologically relevant enzyme design, potentially surpassing structure-based and binding-focused models in creating enzymes capable of catalyzing specific reactions. We provide \textsc{GENzyme} code at \url{https://github.com/WillHua127/GENzyme}.
\end{abstract}

\begin{figure*}[ht!]
\vspace{-0.3cm}
\centering
{
\includegraphics[width=1\textwidth]{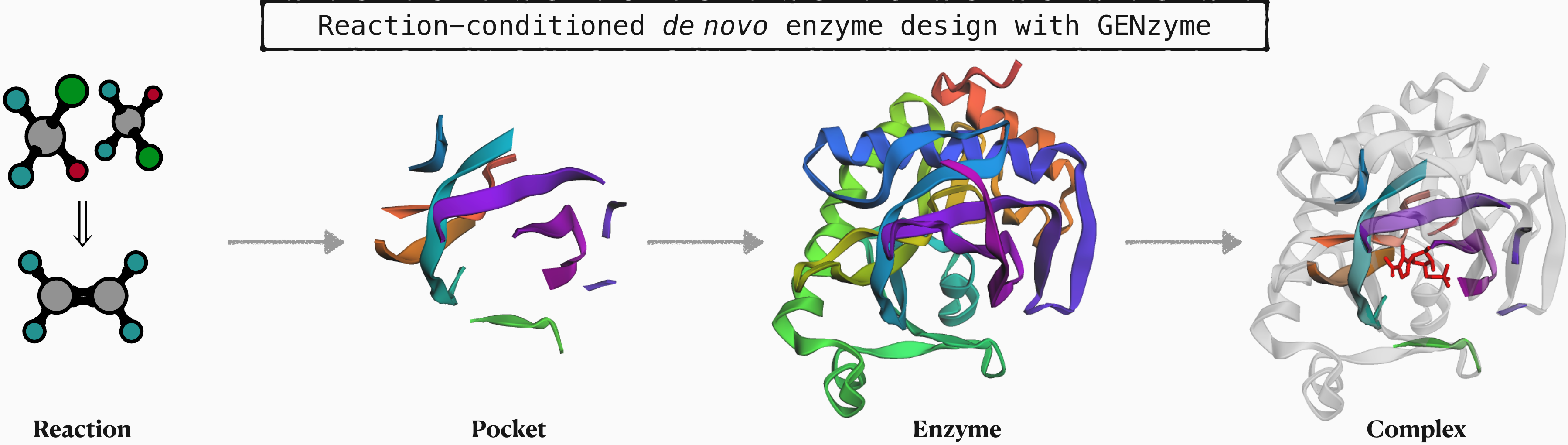}}
{
  \vspace{-0.7cm}
}
\end{figure*}

\section{Introduction}
\begin{figure*}[t!]
\centering
{
\vspace{-0.3cm}
\includegraphics[width=1\textwidth]{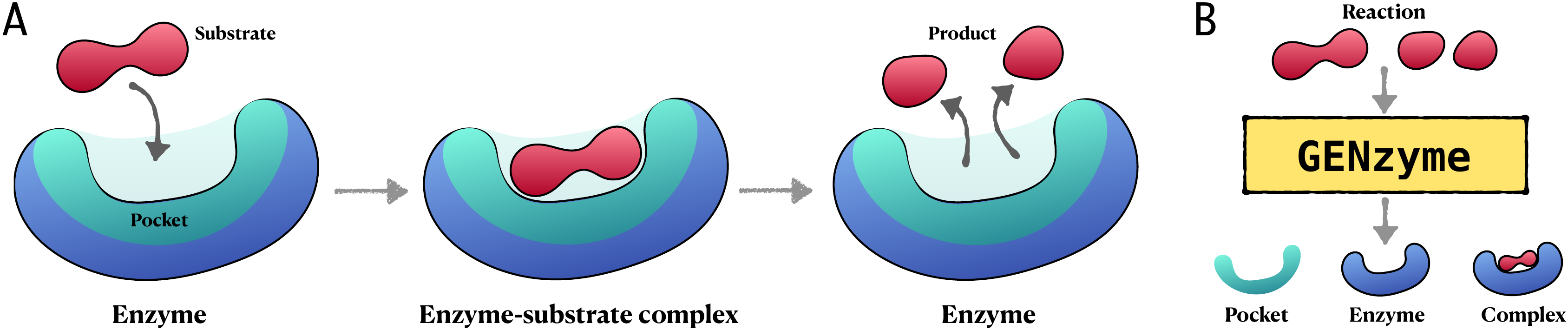}}
{
\vspace{-0.5cm}
  \caption{(A) Enzyme-substrate interaction mechanism. The substrate molecule binds to the enzyme catalytic pocket, where it undergoes catalysis and is converted into product molecules.   (B) The overarching goal of \textsc{GENzyme} for \textit{de novo} enzyme design. Starting from a catalytic reaction, it generates the catalytic pocket, complete enzyme structure, and the enzyme-substrate complex.}
  \label{fig:mechanism}
  \vspace{-0.2cm}
}
\end{figure*}

\vspace{-0.1cm}
\footnotesize
\begin{center}
    \textit{``Congrats to \textbf{David Baker, Demis Hassabis, and John Jumper} for winning the 2024 Nobel Prize in Chemistry. \\ Our research on AI-driven protein design would not have been possible without the groundbreaking contributions of their foundational work. We feel grateful for everything."} — by Authors
\end{center}
\normalsize
Proteins are fundamental to life, playing an important role in numerous biological processes through their involvement in essential interactions \citep{whitford2013proteins, nam2024perspectives}. 
Among these, enzymes stand out as a specialized class of proteins that function as catalysts, driving and regulating nearly all chemical reactions and metabolic pathways in living organisms, from simple bacteria to complex mammals \citep{kraut1988enzymes, murakami1996artificial, copeland2023enzymes} (illustrated in Fig.~\ref{fig:mechanism}(A)). 
The catalytic efficiency of enzymes is central to biological functions, facilitating the rapid production of complex organic molecules necessary for biosynthesis \citep{liu2007cofactor, ferrer2008structure, reetz2024engineered} and enabling the creation of novel biological pathways in synthetic biology \citep{keasling2010manufacturing, hodgman2012cell, girvan2016applications}. Studying enzyme functions across diverse species provides insight into the evolutionary mechanisms that shape metabolic networks, allowing organisms to adapt to their environments \citep{jensen1976enzyme, glasner2006evolution, campbell2016role, pinto2022exploiting}. Enzymes are integral to cellular functions, from energy production to the regulation of genetic information \citep{babcock1992oxygen, heinrich2012regulation, nielsen2016engineering}. Understanding enzyme catalysis, mechanisms, and how enzymes interact with their substrates is not only fundamental to enhance our knowledge of core biological processes but also important for advancing biotechnology and developing new therapeutic approaches to disease management \citep{nam2024perspectives, bell2024strategies, listov2024opportunities}.

While traditional approaches in enzyme research have focused on predicting enzyme functions, annotating enzyme-reaction relationships \citep{gligorijevic2021structure, yu2023enzyme}, or retrieving enzyme-reaction pairs \citep{mikhael2024clipzyme, hua2024reactzyme, yang2024care}, these methods are limited in their ability to design new enzymes capable of catalyzing specific biological reactions \citep{kroll2023general}. Recent progress in designing enzymes for particular enzyme commission (EC) classes \citep{munsamy2022zymctrl, yang2024conditional} has shown success, with models generating enzyme sequences that closely resemble reference sequences and achieve desired EC classifications. Additionally, \cite{hossack2023building} introduced a method for designing novel enzymes by assembling active site and scaffold libraries, followed by refinement algorithms.

However, these advancements are not without limitations. Designing enzymes based solely on EC classifications can constrain model ability to generalize to novel, unseen reactions. In response, recent models have sought to move away from the EC system, instead aiming to directly comprehend the relationships between enzyme structures and their substrate molecules \citep{mikhael2024clipzyme, yang2024care, hua2024reactzyme}. Another significant challenge is that existing enzyme sequence design models are not yet equipped to comprehend the enzyme-substrate catalytic mechanism. Even when new enzyme sequences fold correctly into 3D structures, the catalytic pocket and the intricate binding interactions between enzymes and substrates are frequently underexplored or unclear.

In this work, we introduce \textsc{GENzyme}, a framework designed to address these challenges by generating enzymes capable of catalyzing previously unseen reactions (illustrated in Fig.~\ref{fig:mechanism}(B)). Moreover, \textsc{GENzyme} aims to address the key question of how enzymes and substrates interact by generating enzyme-substrate binding structures. By leveraging generative models for enzyme scaffolds, active sites, and protein language models trained for enzyme motif scaffolding, \textsc{GENzyme} takes a catalytic reaction as input and generates the catalytic pocket, full enzyme structure, and enzyme-substrate complex.
\textbf{\textsc{GENzyme} seeks to advance \textit{de novo} enzyme design for specific catalytic reactions, enhance our understanding of enzyme active sites, and reveal how enzymes interact with substrates during catalysis—ultimately contributing to the understanding of metabolic pathways and potentially aiding therapeutic interventions and disease management.}

\vspace{-0.1cm}
\section{Related Work—Nobel-winning Protein Design Models}
\vspace{-0.1cm}
The field of AI-driven protein design was largely propelled by the success of AlphaFold \citep{senior2020improved}, which achieved remarkable accuracy in protein structure prediction. AlphaFold2 \citep{jumper2021highly} further revolutionized the field, reaching experimental-level folding accuracy and solving the $50$-year-old protein folding problem. This breakthrough provided millions of valid protein structure templates for nearly all catalogued proteins, a feat unattainable for decades. Similarly, RoseTTAFold \citep{baek2021accurate} achieved high folding accuracy, comparable to AlphaFold2. These accomplishments were recognized with the Nobel Prize in Chemistry in 2024.
OpenFold \citep{ahdritz2024openfold}, an open-source project inspired by AlphaFold2, aimed to make protein folding more accessible and transparent. ESMFold \citep{lin2022language} applied ESM-series protein language models trained on millions of sequences, leveraging evolutionary information to predict protein structures at scale. ColabFold \citep{mirdita2022colabfold} streamlined the folding process by integrating AlphaFold2 and RoseTTAFold with MSAs, making protein structure prediction faster and easier via a web server.

\begin{figure*}[t!]
\centering
{
\vspace{-0.3cm}
\includegraphics[width=1\textwidth]{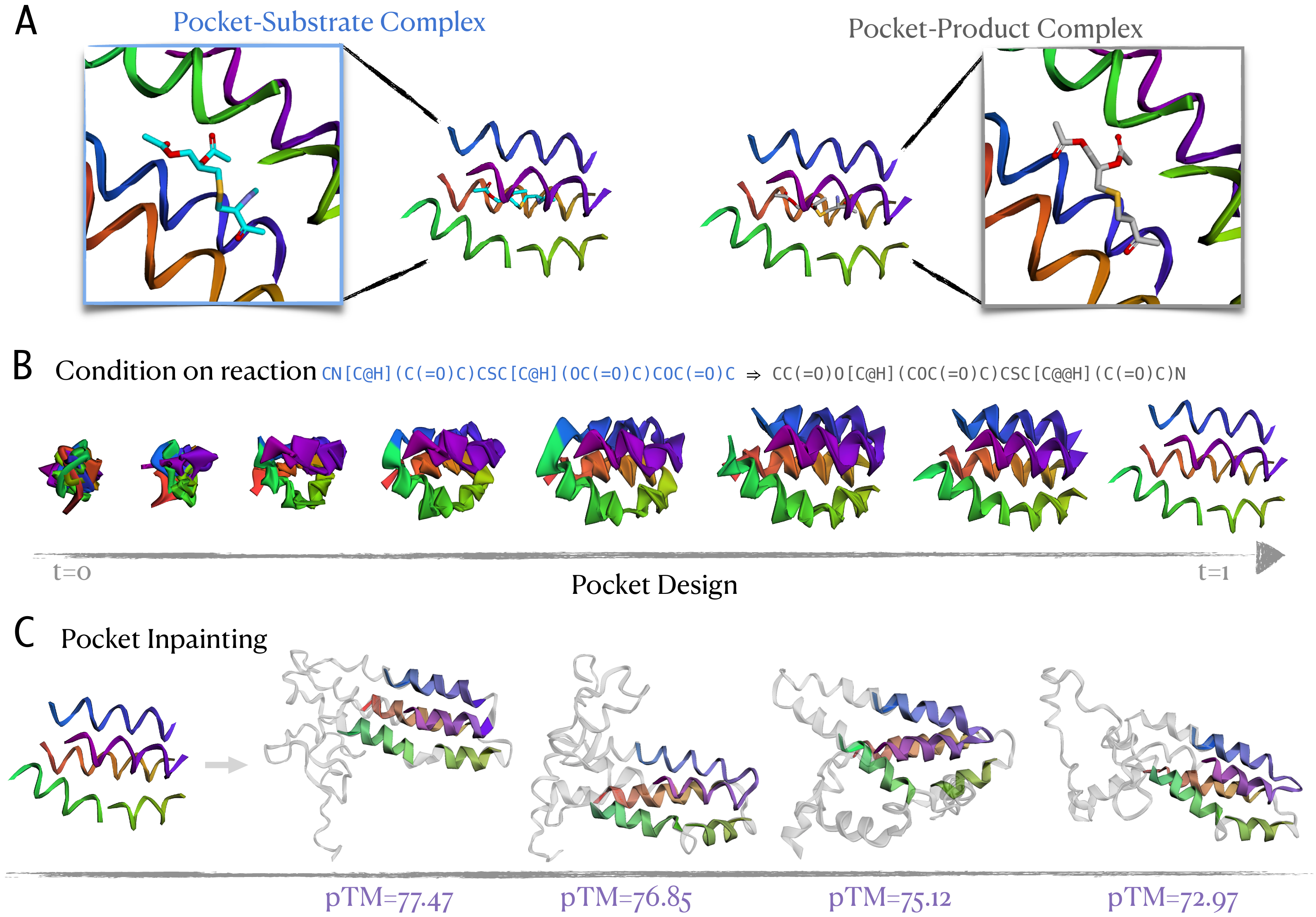}}
{
\vspace{-0.5cm}
  \caption{\textbf{\textit{De novo} Enzyme Design with \textsc{GENzyme}}. (A) Pocket-substrate and pocket-product complexes generated by \textsc{GENzyme}, with the \textcolor{cyan}{substrate molecule shown in cyan} and the \textcolor{gray}{product molecule shown in gray}. The positions of catalytic regions, substrate, and product conformations remain mostly unchanged after catalysis. (B) Reaction-conditioned iterative catalytic pocket generation with \textsc{GENzyme}. The catalytic pocket at \(t=1\) (consists of \(64\) residues) is generated progressively from SE(3) noise initialized at \(t=0\). Additional visualizations are available in App.~\ref{app:alphaenzyme.pocket.design}. (C) Catalytic pocket inpainting with \textsc{GENzyme}, where the generated enzymes achieve high \texttt{pTM} scores, indicating structural quality and alignment with desired functions.}
  \label{fig:alphaenzyme_pocket}
  \vspace{-0.3cm}
}
\end{figure*}

These advancements in protein folding laid the foundation for structure-based protein design models. RFDiffusion \citep{watson2023novo} showcased the ability to generate novel protein structures—ranging from monomers to oligomers and binders—using a generative diffusion approach \citep{song2020score}. Other models, such as Genie \citep{lin2023generating}, Chroma \citep{ingraham2023illuminating}, FoldingDiff \citep{wu2024protein}, FrameDiff \citep{yim2023se}, FoldFlow \citep{bose2023se}, and EvoDiff \citep{alamdari2023protein}, applied various diffusion and flow-matching \citep{lipman2022flow} strategies to design new proteins from random noise, demonstrating good novelty, diversity, and designability.

Beyond monomer designs, recent models have shifted toward designing proteins and their potential binders, including other proteins, molecules, antibodies, RNAs, and others. AlphaFold-Multimer \citep{evans2021protein} extended the folding models to predict protein complexes, learning how proteins interact with each other. RoseTTAFoldNA \citep{baek2024accurate} was developed to predict 3D structures of protein-DNA and protein-RNA complexes, as well as RNA tertiary structures, capturing the interactions between proteins and nucleic acids. RFDiffusionAA \citep{krishna2024generalized} enhanced RFDiffusion by incorporating all-atom representations, allowing it to generate detailed protein structures and ligand interactions. AlphaFold3 \citep{abramson2024accurate} advanced AlphaFold2 by predicting not only individual protein structures but also their interactions with DNA, RNA, and small molecules. Chai1 \citep{Chai-1-Technical-Report}, inspired by AlphaFold3, offers an open-source web server for predicting protein interactions with other biological molecules, showing competitive results. AlphaProteo \citep{zambaldi2024novo} and BindCraft \citep{pacesa2024bindcraft} focus on generating high-strength protein binders that can tightly bind to target molecules, such as viral or cancer proteins.

\begin{figure*}[t!]
\centering
{
\vspace{-0.3cm}
\includegraphics[width=1\textwidth]{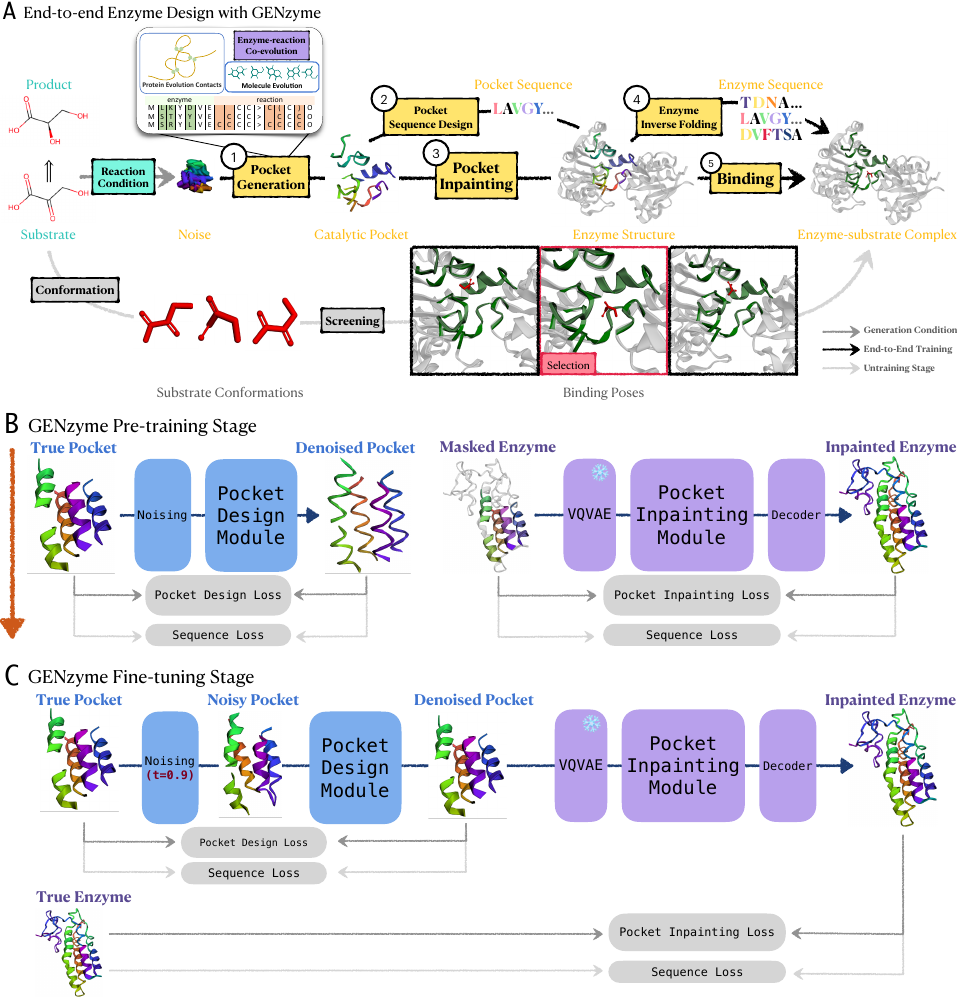}}
{
\vspace{-0.5cm}
  \caption{(A) \textbf{Reaction-conditioned enzyme design with \textsc{GENzyme}}. In example, the design process is conditioned on reaction \footnotesize\texttt{OCC(=O)C(=O)O $\Rightarrow$ OC[C@H](C(=O)O)O}\normalsize. \textsc{GENzyme} designs enzymes by first {\textcircled{\raisebox{-0.9pt}{1}}} generating catalytic pockets, and {\textcircled{\raisebox{-0.9pt}{2}}} co-designing pocket sequence. Next, it {\textcircled{\raisebox{-0.9pt}{3}}} inpaints the catalytic pocket to complete the full enzyme structure with {\textcircled{\raisebox{-0.9pt}{4}}} the generation of enzyme sequence. Finally, {\textcircled{\raisebox{-0.9pt}{5}}} the substrate conformation binds to the catalytic pocket of the full enzyme, and the ideal lock-and-key enzyme-substrate complex is predicted. (B) \textbf{\textsc{GENzyme} Pre-training Phase}, in which each module is trained individually on catalytic pockets and full enzyme structures. (C) \textbf{\textsc{GENzyme} Fine-tuning Phase}, where the model undergoes end-to-end training with slight perturbations applied to input pockets and enzymes.}
  \label{fig:alphaenzyme}
  \vspace{-0.2cm}
}
\end{figure*}

While structure-based protein design models are highly effective for creating stable and functional proteins, their primary focus is on static protein-ligand or protein-protein interactions. Enzymes, as catalysts that regulate nearly all chemical reactions and metabolic pathways in the human body, present a unique challenge for these models. Enzyme-substrate interactions are often driven by dynamic energy changes, and the induced-fit model is typically sufficient for catalysis. During a catalytic reaction, the enzyme transforms substrates into products through a chemical process, and afterward, it is free to bind another substrate molecule. The dynamic change and chemical transformation are completely different to static mechanism in protein-ligand interactions.

\begin{figure*}[t!]
\centering
{
\vspace{-0.3cm}
\includegraphics[width=1\textwidth]{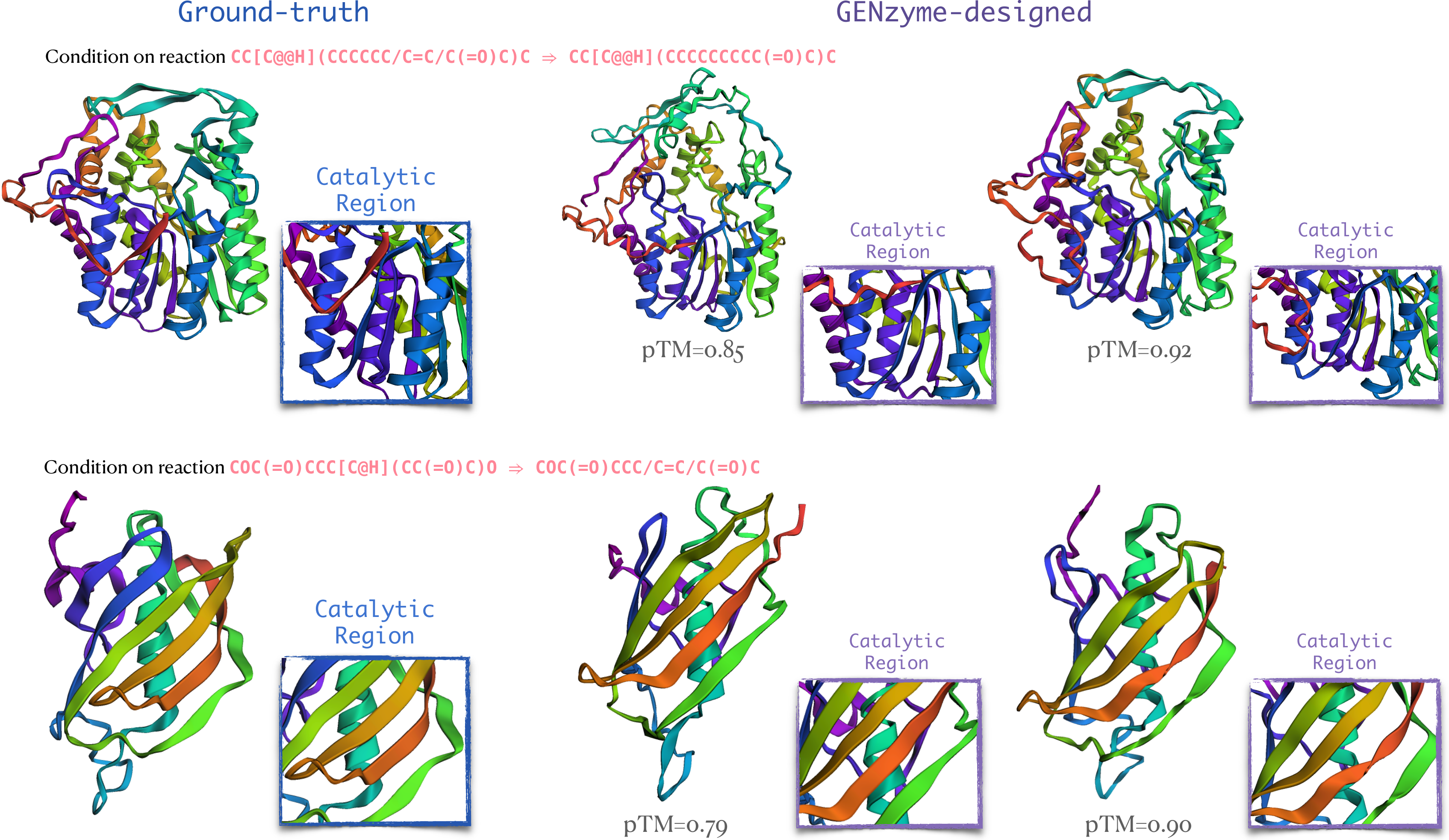}}
{
\vspace{-0.5cm}
  \caption{\textsc{GENzyme} \textit{de novo} Enzyme Design Example.}
  \label{fig:visual.example}
  \vspace{-0.3cm}
}
\end{figure*}

As a result, structure-based design models, which focus on static ligand-binding interactions, may fail to capture the dynamic transformations and complexities of enzyme-substrate interactions. In contrast, \textcolor{RoyalBlue}{\textsc{GENzyme} proposes a function-based approach to enzyme design, where enzyme function is defined by the chemical reaction it catalyzes, potentially addressing the limitations of current structure-based models and allowing for the design of enzymes for specific catalytic functions.}

Given any catalytic reaction \(m_r\), \textsc{GENzyme} aims to generate/predict the enzyme \(\mathbf{E}\), which is capable of potentially catalyzing \(m_r\), and the corresponding enzyme-substrate complex \(\mathbf{C}\), as:
\small
\begin{equation}
   \textcolor{RoyalBlue}{\textsc{Pocket } \mathbf{E^P}, \textsc{Enzyme } \mathbf{E}, \textsc{Complex } \mathbf{C} \leftarrow \textsc{GENzyme}(\textsc{Reaction } m_r)}.
\end{equation}
\normalsize
Fig.~\ref{fig:alphaenzyme} illustrates the process by which \textsc{GENzyme} generates an enzyme and its corresponding docked enzyme-substrate complex. Starting with the input SMILES representations of the reaction, \textsc{GENzyme} generates the full enzyme structure and sequence from SE(3) noise, progressing through parts 1 to 4, where the catalytic pocket is first designed and then inpainted into a complete enzyme representation. Concurrently, the binding model computes multiple substrate conformations, optimizing both the geometry and binding poses based on the catalytic pocket of the full enzyme. An optimal enzyme-substrate model is predicted and output, as illustrated in part 5 of Fig.~\ref{fig:alphaenzyme}.

\vspace{-0.1cm}
\section{GENzyme Result}
\vspace{-0.1cm}
We evaluate \textsc{GENzyme} from both structural and functional perspectives, starting with an assessment of the quality of the generated catalytic pockets, followed by an evaluation of the inpainted full enzyme structures. \textsc{GENzyme} is compared against several baselines for enzyme design, including EnzymeFlow (pocket-level) \citep{hua2024enzymeflow}, ZymCTRL+ESMFold (enzyme-level) \citep{munsamy2022zymctrl, lin2022language}, and RFDiffusionAA (pocket- and enzyme-level) \citep{krishna2024generalized}. Additionally, we employ Chai1 \citep{Chai-1-Technical-Report} to assess the quality of enzyme-substrate complexes. 
For catalytic pockets and enzymes generated by RFDiffusionAA, we apply LigandMPNN \citep{dauparas2023atomic} to inverse fold and predict sequences post-hoc.

For each model, we generate $8$\footnote{We limit sample generation to $8$ due to wet-lab experimental constraints, as producing excessive samples is cost-prohibitive for laboratory trials. Typically, $3$, $5$, $8$, or $10$ samples are adequate for such trials.} catalytic pockets and full enzyme structures per reaction (or substrate for RFDiffusionAA, EC-class for ZymCTRL) to comprehensively evaluate their performance. For \textsc{GENzyme}, we randomly generate and select $8$ enzymes with $\texttt{pTM}>0.60$ or $\texttt{pLDDT}>0.60$\footnote{We use a linear schedular of \texttt{pTM} and \texttt{pLDDT} for filtering, starting with $\texttt{pTM}>0.80$ or $\texttt{pLDDT}>0.80$.} and use their corresponding catalytic pockets for evaluation. Both structural validity and catalytic functionality are assessed for the generated pockets and enzymes.
We provide the code of \textsc{GENzyme} and inference scripts at \url{https://github.com/WillHua127/GENzyme}.

\footnotesize
\begin{table}[t!]
  \centering
  \vspace{-0.3cm}
  \caption{Evaluation data for enzyme design, featuring six enzyme-reaction pairs from different enzyme commission classes, representing a variety of functions, catalysis, and reaction types.}
  \vspace{0.1cm}
  \resizebox{1\textwidth}{!}{
    \begin{tabular}{l|l|l}
    \toprule
    UniProt & Substrate & Product \\
    \midrule
    \rowcolor[rgb]{ .855,  .914,  .973} Q8N4T8 & \texttt{CC(=O)CC(=O)CC(=O)OC} & \texttt{CC(=O)C[C@@H](CC(=O)OC)O} \\
    \rowcolor[rgb]{ .855,  .914,  .973} B2SUY7 & \texttt{CC[C@@H](CCCCCC/C=C/C(=O)C)C} & \texttt{CC[C@@H](CCCCCCCCC(=O)C)C} \\
    \rowcolor[rgb]{ .855,  .914,  .973} Q9WYD4 & \texttt{OC[C@H]([C@H]([C@@H](C(=O)CO)O)O)O} & \texttt{OC[C@H]([C@H]([C@@H]([C@@H](CO)O)O)O)O} \\
    \rowcolor[rgb]{ .855,  .914,  .973} Q86W10 & \texttt{CN1c2cc(C)c(cc2Nc2c1[nH]c(=O)[nH]c2=O)C} & \texttt{Cc1cc2nc3-c(n(c2cc1C)C)nc(=O)[nH]c3=O} \\
    \rowcolor[rgb]{ .855,  .914,  .973} Q2TZB2 & \texttt{N[C@H](C(=O)O)CCCN} & \texttt{ONCCC[C@@H](C(=O)O)N} \\
    \midrule
    \rowcolor[rgb]{ .753,  .902,  .961} Q4V8E6 & \texttt{SC[C@@H](C(=O)NCC(=O)O)NC(=O)CC[C@@H](C(=O)O)N} & \texttt{CSC[C@@H](C(=O)NCC(=O)O)NC(=O)CC[C@@H](C(=O)O)N} \\
    \rowcolor[rgb]{ .753,  .902,  .961} A6UP94 & \texttt{CC(=O)c1ccccc1} & \texttt{C[C@H](c1ccccc1)N} \\
    \rowcolor[rgb]{ .753,  .902,  .961} Q8DBS9 & \texttt{OC[C@H]1O[C@@H](C[C@@H]1O)OP(=O)(O)O} & \texttt{OC[C@H]1O[C@H](C[C@@H]1O)n1cnc2c1ncnc2N} \\
    \rowcolor[rgb]{ .753,  .902,  .961} A1JJD0 & \texttt{OC[C@@]1(O)O[C@@H]([C@H]([C@@H]1O)O)COP(=O)(O)O} & \texttt{O=C[C@@H](COP(=O)(O)O)O} \\
    \rowcolor[rgb]{ .753,  .902,  .961} C4JE77 & \texttt{C[S@+](C[C@H]1O[C@H]([C@@H]([C@@H]1O)O)n1cnc2c1ncnc2N)CC[C@@H](C(=O)O)N} & \texttt{C[C@H]1O[C@H]([C@@H]([C@@H]1O)O)n1cnc2c1ncnc2N} \\
    \midrule
    \rowcolor[rgb]{ .984,  .886,  .835} O84413 & \texttt{CN[C@H](C(=O)C)CSC[C@H](OC(=O)C)COC(=O)C} & \texttt{CC(=O)O[C@H](COC(=O)C)CSC[C@@H](C(=O)C)N} \\
    \rowcolor[rgb]{ .984,  .886,  .835} B2HDR8 & \texttt{O=CN(c1ccc(cc1)C(=O)N[C@H](C(=O)O)CCC(=O)C)C[C@H]1CNc2c(N1)c(=O)[nH]c(n2)N} & \texttt{CC(=O)CC[C@@H](C(=O)O)NC(=O)c1ccc(cc1)N1C=[N+]2[C@@H](C1)CNc1c2c(=O)[nH]c(n1)N} \\
    \rowcolor[rgb]{ .984,  .886,  .835} F1RT67 & \texttt{O[C@@H]1[C@H](OP(=O)(O)O)[C@H](O[C@H]1n1cnc2c1ncnc2N)COP(=O)(O)O} & \texttt{O[C@@H]1[C@H](O)[C@H](O[C@H]1n1cnc2c1ncnc2N)COP(=O)(O)O} \\
    \rowcolor[rgb]{ .984,  .886,  .835} B3LNM2 & \texttt{CC(=O)[C@@H](NC(=O)[C@H](CCSC)N)C} & \texttt{CSCC[C@@H](C(=O)O)N} \\
    \rowcolor[rgb]{ .984,  .886,  .835} P0AA00 & \texttt{N[C@H](c1ccccc1)C(=O)N[C@@H]1C(=O)N2[C@@H]1SC([C@@H]2C(=O)O)(C)C} & \texttt{N[C@H](c1ccccc1)C(=O)N[C@H](C1N[C@H](C(S1)(C)C)C(=O)O)C(=O)O} \\
    \midrule
    \rowcolor[rgb]{ .757,  .941,  .784} P61452 & \texttt{O[C@H](C(C)C)CC(=O)C} & \texttt{CC(/C=C/C(=O)C)C} \\
    \rowcolor[rgb]{ .757,  .941,  .784} A5N2N5 & \texttt{OC(=O)C(=C)OP(=O)(O)O} & \texttt{OC[C@H](C(=O)O)OP(=O)(O)O} \\
    \rowcolor[rgb]{ .757,  .941,  .784} C6DGZ5 & \texttt{O[C@@H]([C@H](c1c[nH]c2c1cccc2)O)COP(=O)(O)O} & \texttt{OC(=O)[C@H](Cc1c[nH]c2c1cccc2)N} \\
    \rowcolor[rgb]{ .757,  .941,  .784} Q9RVF5 & \texttt{COC(=O)CCC[C@H](CC(=O)C)O} & \texttt{COC(=O)CCC/C=C/C(=O)C} \\
    \rowcolor[rgb]{ .757,  .941,  .784} A7MKL4 & \texttt{O[C@@H]([C@H]([C@@H](CC(=O)C(=O)O)O)O)COP(=O)(O)O} & \texttt{OP(=O)(O)O} \\
    \midrule
    \rowcolor[rgb]{ .949,  .808,  .937} B7UVB1 & \texttt{CCCCCCCCCCCCCCCCC[C@H](CC(=O)C)O} & \texttt{CCCCCCCCCCCCCCCCC/C=C/C(=O)C} \\
    \rowcolor[rgb]{ .949,  .808,  .937} Q6DDT1 & \texttt{OC1O[C@H](COP(=O)(O)O)[C@H]([C@@H]([C@H]1O)O)O} & \texttt{O[C@@H]1[C@H](OP(=O)(O)O)[C@@H](O)[C@@H]([C@H]([C@@H]1O)O)O} \\
    \rowcolor[rgb]{ .949,  .808,  .937} P9WIQ1 & \texttt{OC[C@H]1O[C@H](OP(=O)(OP(=O)(OC[C@H]2O[C@H]([C@@H]([C@@H]2O)O)n2ccc(=O)[nH]c2=O)O)O)[C@@H]([C@H]([C@H]1O)O)O} & \texttt{OC[C@H]([C@@H]1O[C@@H]([C@@H]([C@H]1O)O)OP(=O)(OP(=O)(OC[C@H]1O[C@H]([C@@H]([C@@H]1O)O)n1ccc(=O)[nH]c1=O)O)O)O} \\
    \rowcolor[rgb]{ .949,  .808,  .937} Q6CFH4 & \texttt{OCC(=O)[C@@H]([C@@H](COP(=O)(O)O)O)O} & \texttt{O=C[C@@H]([C@@H]([C@@H](COP(=O)(O)O)O)O)O} \\
    \rowcolor[rgb]{ .949,  .808,  .937} Q23381 & \texttt{O=C(NCCSC(=O)[C@@H](C(=O)O)C)CCNC(=O)[C@@H](C(COP(=O)(OP(=O)(OC[C@H]1O[C@H]([C@@H]([C@@H]1OP(=O)(O)O)O)n1cnc2c1ncnc2N)O)O)(C)C)O} & \texttt{O=C(NCCSC(=O)CCC(=O)O)CCNC(=O)[C@@H](C(COP(=O)(OP(=O)(OC[C@H]1O[C@H]([C@@H]([C@@H]1OP(=O)(O)O)O)n1cnc2c1ncnc2N)O)O)(C)C)O} \\
    \midrule
    \rowcolor[rgb]{ .855,  .949,  .816} B3WC77 & \texttt{C[C@H](C(=O)O)N} & \texttt{CC(=O)[C@H](N)C} \\
    \rowcolor[rgb]{ .855,  .949,  .816} Q8E0G6 & \texttt{C[C@H](C(=O)O)N} & \texttt{C[C@H](C(=O)O)NC(=O)[C@H](N)C} \\
    \rowcolor[rgb]{ .855,  .949,  .816} A9CB42 & \texttt{COP(=O)(OC[C@H]1O[C@H]([C@@H]([C@@H]1OP(=O)(O)O)O)C)O} & \texttt{COP(=O)(OC[C@H]1O[C@H]([C@@H]([C@@H]1OP(=O)(OC[C@H]1O[C@H]([C@@H]([C@@H]1OP(=O)(OC)O)O)C)O)O)C)O} \\
    \rowcolor[rgb]{ .855,  .949,  .816} Q6G456 & \texttt{OCC([C@H](C(=O)O)O)(C)C} & \texttt{OCC([C@H](C(=O)NCCC(=O)O)O)(C)C} \\
    \rowcolor[rgb]{ .855,  .949,  .816} Q0SM64 & \texttt{COP(=O)(O[C@@H]1[C@@H](COP(=O)(OC)O)O[C@H]([C@@H]1O)n1ccc(nc1=O)N)O} & \texttt{COP(=O)(O[C@@H]1[C@@H](COP(=O)(OC)O)O[C@H]([C@@H]1O)n1ccc(=N)nc1NCCCC[C@@H](C(=O)O)N)O} \\
    \bottomrule
    \end{tabular}%
    }
  \label{tab:eval.data}%
  \vspace{-0.2cm}
\end{table}%
\normalsize

\textbf{Evaluation Data.}
For model validation, we select five enzyme-reaction pairs from each EC class (1-6) in the EnzymeFill dataset \citep{hua2024enzymeflow}, capturing diverse enzyme functions, catalytic mechanisms, and reaction types \citep{schomburg2002brenda}. Using MMseqs2 \citep{steinegger2017mmseqs2}, we cluster at a $10\%$ homology threshold and select the central member of each cluster as the initial dataset. After removing duplicates of repeated reactions and UniProt, we uniformly sample $5$ enzyme-reaction pairs per EC class for evaluation, yielding a total of $30$ unique catalytic pockets and $30$ unique reactions. The full set of evaluation enzyme-reaction pairs is presented in Tab.~\ref{tab:eval.data}.

\vspace{-0.1cm}
\subsection{Enzyme Structure Evaluation}
\vspace{-0.1cm}
Following \cite{hua2024enzymeflow}, we begin by evaluating the structural validity of the generated catalytic pockets and full enzymes. While enzyme function determines whether the designed pocket or enzyme can catalyze a specific reaction, structure provides important information regarding substrate binding. Proper substrate binding to the active region of the enzyme is essential for catalysis and chemical reaction, even if a perfect binding pose is not always needed.

To assess structural validity, we use the metrics:
\textbf{Root Mean Square Deviation (\texttt{RMSD})}: Measures the structural distance between ground-truth and generated catalytic pockets and enzymes, indicating the alignment accuracy of the generated structures with the actual structures.
\textbf{\texttt{TM-score}}: Assesses topological similarity between generated and ground-truth structures, particularly focusing on local deviations \citep{zhang2005tm}.
\textbf{Embedding Distance (\texttt{Emb-Dist})}: Calculates the embedding distance between generated and ground-truth pockets and enzymes. We co-encode both protein sequences and structures using ESM3 \citep{hayes2024simulating}, and employ tSNE to reduce hidden dimensionality to compute embedding distances.

\begin{table}[ht!]
  \vspace{-0.5cm}
  \centering
  \caption{Evaluation of structural validity of \textsc{GENzyme}- and baseline-generated catalytic pockets. We highlight the top performing results in \textbf{bold}.}
  \vspace{0.1cm}
  \resizebox{1.\columnwidth}{!}{%
    \begin{tabular}{c|c|ccc|ccc|ccc}
    \toprule
    \multicolumn{2}{c|}{\multirow{2}[4]{*}{Pocket Evaluation}} & \multicolumn{3}{c|}{\texttt{RMSD} ($\downarrow$)} & \multicolumn{3}{c|}{\texttt{TM-score}($\uparrow$)} & \multicolumn{3}{c}{\texttt{Emb-Dist}($\downarrow$)}\\
\cmidrule{3-11}    \multicolumn{2}{c|}{} & \texttt{Top1}  & \texttt{Top3}  & \texttt{Mean}  & \texttt{Top1}  & \texttt{Top3}  & \texttt{Mean}  & \texttt{Top1}  & \texttt{Top3}  & \texttt{Mean}\\
    \midrule
    \multirow{3}[2]{*}{EC1} & \cellcolor[rgb]{ .855,  .949,  .816}EnzymeFlow & 3.76  & 3.90  & 4.19  & 0.29  & 0.28  & 0.26  & 19.82 & 20.29 & 21.48\\
          & \cellcolor[rgb]{ 1,  1,  0}RFDiffusionAA & 3.71  & 3.82  & 4.18  & 0.28  & 0.27  & 0.25  & 7.33  & 8.05  & 9.28\\
          & \cellcolor[rgb]{ .792,  .929,  .984}\textsc{GENzyme} & \textbf{1.88} & \textbf{2.09} & \textbf{2.41} & \textbf{0.67} & \textbf{0.65} & \textbf{0.59} & \textbf{0.25} & \textbf{0.58} & \textbf{1.54}\\
    \midrule
    \multirow{3}[2]{*}{EC2} & \cellcolor[rgb]{ .855,  .949,  .816}EnzymeFlow & 3.78  & 3.93  & 4.19  & 0.28  & 0.27  & 0.25  & 43.30 & 44.21 & 47.15 \\
          & \cellcolor[rgb]{ 1,  1,  0}RFDiffusionAA & 3.47  & \textbf{3.78} & 4.17  & 0.27  & 0.27  & 0.25  & 26.91 & 27.84 & 30.44\\
          & \cellcolor[rgb]{ .792,  .929,  .984}\textsc{GENzyme} & \textbf{3.65} & 3.80  & \textbf{4.09} & \textbf{0.33} & \textbf{0.31} & \textbf{0.28} & \textbf{1.03} & \textbf{1.85} & \textbf{4.40}\\
    \midrule
    \multirow{3}[2]{*}{EC3} & \cellcolor[rgb]{ .855,  .949,  .816}EnzymeFlow & 3.46  & 3.69  & 4.12  & 0.31  & 0.29  & 0.27  & 9.84  & 10.02 & 11.14\\
          & \cellcolor[rgb]{ 1,  1,  0}RFDiffusionAA & 3.25  & 3.50  & 3.96  & 0.31  & 0.30  & 0.27  & 10.83 & 10.96 & 12.22\\
          & \cellcolor[rgb]{ .792,  .929,  .984}\textsc{GENzyme} & \textbf{3.18} & \textbf{3.32} & \textbf{3.84} & \textbf{0.39} & \textbf{0.36} & \textbf{0.32} & \textbf{0.42} & \textbf{1.08} & \textbf{2.19}\\
    \midrule
    \multirow{3}[2]{*}{EC4} & \cellcolor[rgb]{ .855,  .949,  .816}EnzymeFlow & 3.31  & 3.48  & 3.91  & 0.32  & 0.30  & 0.27  & 18.02 & 18.56 & 19.96\\
          & \cellcolor[rgb]{ 1,  1,  0}RFDiffusionAA & 3.00  & 3.29  & 3.79  & 0.30  & 0.29  & 0.27  & 14.62 & 15.62 & 17.39\\
          & \cellcolor[rgb]{ .792,  .929,  .984}\textsc{GENzyme} & \textbf{2.58} & \textbf{2.73} & \textbf{2.96} & \textbf{0.56} & \textbf{0.54} & \textbf{0.51} & \textbf{0.50} & \textbf{1.07} & \textbf{2.43}\\
    \midrule
    \multirow{3}[2]{*}{EC5} & \cellcolor[rgb]{ .855,  .949,  .816}EnzymeFlow & 3.48  & 3.66  & 4.06  & 0.29  & 0.28  & 0.26  & 13.71 & 14.01 & 14.81\\
          & \cellcolor[rgb]{ 1,  1,  0}RFDiffusionAA & 3.39  & 3.61  & 3.98  & 0.31  & 0.29  & 0.26  & 7.83  & 8.23  & 8.93\\
          & \cellcolor[rgb]{ .792,  .929,  .984}\textsc{GENzyme} & \textbf{2.87} & \textbf{3.18} & \textbf{3.62} & \textbf{0.44} & \textbf{0.43} & \textbf{0.39} & \textbf{0.74} & \textbf{1.07} & \textbf{1.78}\\
    \midrule
    \multirow{3}[2]{*}{EC6} & \cellcolor[rgb]{ .855,  .949,  .816}EnzymeFlow & 3.60  & 3.79  & 4.13  & 0.31  & 0.30  & 0.28  & 16.00 & 16.56 & 18.19\\
          & \cellcolor[rgb]{ 1,  1,  0}RFDiffusionAA & 3.43  & 3.69  & 4.12  & 0.30  & 0.29  & 0.26  & 19.08 & 19.73 & 21.21\\
          & \cellcolor[rgb]{ .792,  .929,  .984}\textsc{GENzyme} & \textbf{3.14} & \textbf{3.32} & \textbf{3.73} & \textbf{0.43} & \textbf{0.40} & \textbf{0.36} & \textbf{1.23} & \textbf{1.45} & \textbf{2.56}\\
    \bottomrule
    \end{tabular}%
    }
  \label{tab:pocket.eval}%
\end{table}%

\begin{table}[ht!]
\vspace{-0.4cm}
  \centering
  \caption{Evaluation of structural validity of \textsc{GENzyme}-inpainted and baseline-generated enzymes. We highlight the top performing results in \textbf{bold}.}
  \vspace{0.1cm}
  \resizebox{1.\columnwidth}{!}{%
    \begin{tabular}{c|c|ccc|ccc|ccc}
    \toprule
    \multicolumn{2}{c|}{\multirow{2}[4]{*}{Enzyme Evaluation}} & \multicolumn{3}{c|}{\texttt{RMSD} ($\downarrow$)} & \multicolumn{3}{c|}{\texttt{TM-score}($\uparrow$)} & \multicolumn{3}{c}{\texttt{Emb-Dist}($\downarrow$)}\\
\cmidrule{3-11}    \multicolumn{2}{c|}{} & \texttt{Top1}  & \texttt{Top3}  & \texttt{Mean}  & \texttt{Top1}  & \texttt{Top3}  & \texttt{Mean}  & \texttt{Top1}  & \texttt{Top3}  & \texttt{Mean}\\
    \midrule
    \multirow{3}[2]{*}{EC1} & \cellcolor[rgb]{ .91,  .91,  .91}ZymCTRL+ESMFold & \textbf{3.48} & \textbf{4.08} & 5.04  & 0.59  & 0.49  & 0.39  & 2.48  & 5.71  & 10.96\\
          & \cellcolor[rgb]{ 1,  1,  0}RFDiffusionAA & 4.46  & 4.76  & 5.36  & 0.48  & 0.44  & 0.38  & 22.38 & 23.79 & 26.89\\
          & \cellcolor[rgb]{ .792,  .929,  .984}\textsc{GENzyme} & 4.13  & 4.36  & \textbf{4.83} & \textbf{0.62} & \textbf{0.58} & \textbf{0.53} & \textbf{2.41} & \textbf{2.92} & \textbf{5.49} \\
    \midrule
    \multirow{3}[2]{*}{EC2} & \cellcolor[rgb]{ .91,  .91,  .91}ZymCTRL+ESMFold & \textbf{4.74} & \textbf{5.05} & \textbf{5.51} & 0.42  & 0.39  & 0.33  & 18.63 & 19.59 & 21.90\\
          & \cellcolor[rgb]{ 1,  1,  0}RFDiffusionAA & 4.88  & 5.08  & 5.58  & \textbf{0.43} & \textbf{0.42} & \textbf{0.38} & 5.42  & 6.31  & 8.91\\
          & \cellcolor[rgb]{ .792,  .929,  .984}\textsc{GENzyme} & 5.27  & 5.45  & 5.99  & 0.37  & 0.36  & 0.32  & \textbf{5.15} & \textbf{6.25} & \textbf{8.43}\\
    \midrule
    \multirow{3}[2]{*}{EC3} & \cellcolor[rgb]{ .91,  .91,  .91}ZymCTRL+ESMFold & \textbf{4.87} & \textbf{5.02} & \textbf{5.47} & 0.38  & 0.35  & 0.33  & \textbf{1.61} & \textbf{4.65} & \textbf{10.82}\\
          & \cellcolor[rgb]{ 1,  1,  0}RFDiffusionAA & 5.13  & 5.28  & 5.71  & 0.38  & 0.36  & 0.33  & 4.95  & 6.74  & 10.59\\
          & \cellcolor[rgb]{ .792,  .929,  .984}\textsc{GENzyme} & 5.08  & 5.26  & 5.76  & \textbf{0.38} & \textbf{0.38} & \textbf{0.33} & 10.46 & 11.74 & 15.06\\
    \midrule
    \multirow{3}[2]{*}{EC4} & \cellcolor[rgb]{ .91,  .91,  .91}ZymCTRL+ESMFold & \textbf{3.92} & 4.17  & 4.67  & 0.40  & 0.39  & 0.36  & \textbf{2.95} & \textbf{5.98} & {20.43}\\
          & \cellcolor[rgb]{ 1,  1,  0}RFDiffusionAA & 4.49  & 4.68  & 5.25  & 0.45  & 0.42  & 0.37  & 65.60 & 69.49 & 78.31 \\
          & \cellcolor[rgb]{ .792,  .929,  .984}\textsc{GENzyme} & 4.02  & \textbf{4.15} & \textbf{4.59} & \textbf{0.57} & \textbf{0.55} & \textbf{0.52} & 9.61  & 9.85  & \textbf{10.49}\\
    \midrule
    \multirow{3}[2]{*}{EC5} & \cellcolor[rgb]{ .91,  .91,  .91}ZymCTRL+ESMFold & 5.02  & 5.26  & 5.63  & 0.38  & 0.35  & 0.33  & 9.77  & 11.07 & 13.00\\
          & \cellcolor[rgb]{ 1,  1,  0}RFDiffusionAA & \textbf{4.48} & \textbf{4.71} & \textbf{5.22} & \textbf{0.50} & \textbf{0.46} & \textbf{0.40} & \textbf{2.87} & \textbf{3.66} & \textbf{5.31}\\
          & \cellcolor[rgb]{ .792,  .929,  .984}\textsc{GENzyme} & 5.02  & 5.32  & 5.80  & 0.43  & 0.41  & 0.39  & 4.58  & 5.20  & 7.14\\
    \midrule
    \multirow{3}[2]{*}{EC6} & \cellcolor[rgb]{ .91,  .91,  .91}ZymCTRL+ESMFold & 5.25  & 5.46  & 5.75  & 0.37  & 0.35  & 0.33  & \textbf{5.97} & \textbf{7.88} & \textbf{13.25}\\
          & \cellcolor[rgb]{ 1,  1,  0}RFDiffusionAA & \textbf{4.53} & \textbf{4.89} & \textbf{5.50} & \textbf{0.45} & \textbf{0.43} & \textbf{0.38} & 19.38 & 20.76 & 25.87\\
          & \cellcolor[rgb]{ .792,  .929,  .984}\textsc{GENzyme} & 5.76  & 6.00  & 6.42  & 0.34  & 0.32  & 0.28  & 12.24 & 14.73 & 16.87\\
    \bottomrule
    \end{tabular}%
    }
    \vspace{-0.2cm}
  \label{tab:enzyme.eval}%
\end{table}%
We compare the structural validity of \textsc{GENzyme}- and baseline-generated catalytic pockets in Tab.~\ref{tab:pocket.eval}, and the inpainted and generated enzymes in Tab.~\ref{tab:enzyme.eval}, focusing on structural consistency at both local (catalytic regions) and global (entire enzyme) levels.

In Tab.~\ref{tab:pocket.eval}, which evaluates local consistency, \textsc{GENzyme}-generated catalytic pockets demonstrate closer alignment to ground-truth pockets, reflected in lower distance errors and higher structural alignment scores, as well as lower embedding distances. This suggests that \textsc{GENzyme} can design catalytic pockets that maintain or meet specific catalytic requirements for the target reaction. 
In Tab.~\ref{tab:enzyme.eval}, assessing global consistency, we observe that \textsc{GENzyme}-inpainted enzymes differ more from ground-truth structures than the catalytic regions alone, while the embedding distances do not diverge dramatically. This divergence indicates that, while \textsc{GENzyme} preserves critical catalytic properties locally, it also introduces structural modifications across the entire enzyme, potentially enhancing overall stability and functional diversity beyond wild-type enzymes.

These findings are further validated by the t-SNE visualizations in Fig.~\ref{fig:pocket_tsne} and Fig.~\ref{fig:protein_tsne} in Sec.~\ref{sec:tsne.embeddings}, where we analyze embeddings for generated catalytic pockets and full enzymes, supporting the observed balance between local catalytic alignment and broader structural innovation.

\vspace{-0.1cm}
\subsection{Insights from Enzyme Sequence-Structure Embeddings}
\begin{figure*}[ht!]
\vspace{-0.5cm}
\centering
{
\includegraphics[width=1\textwidth]{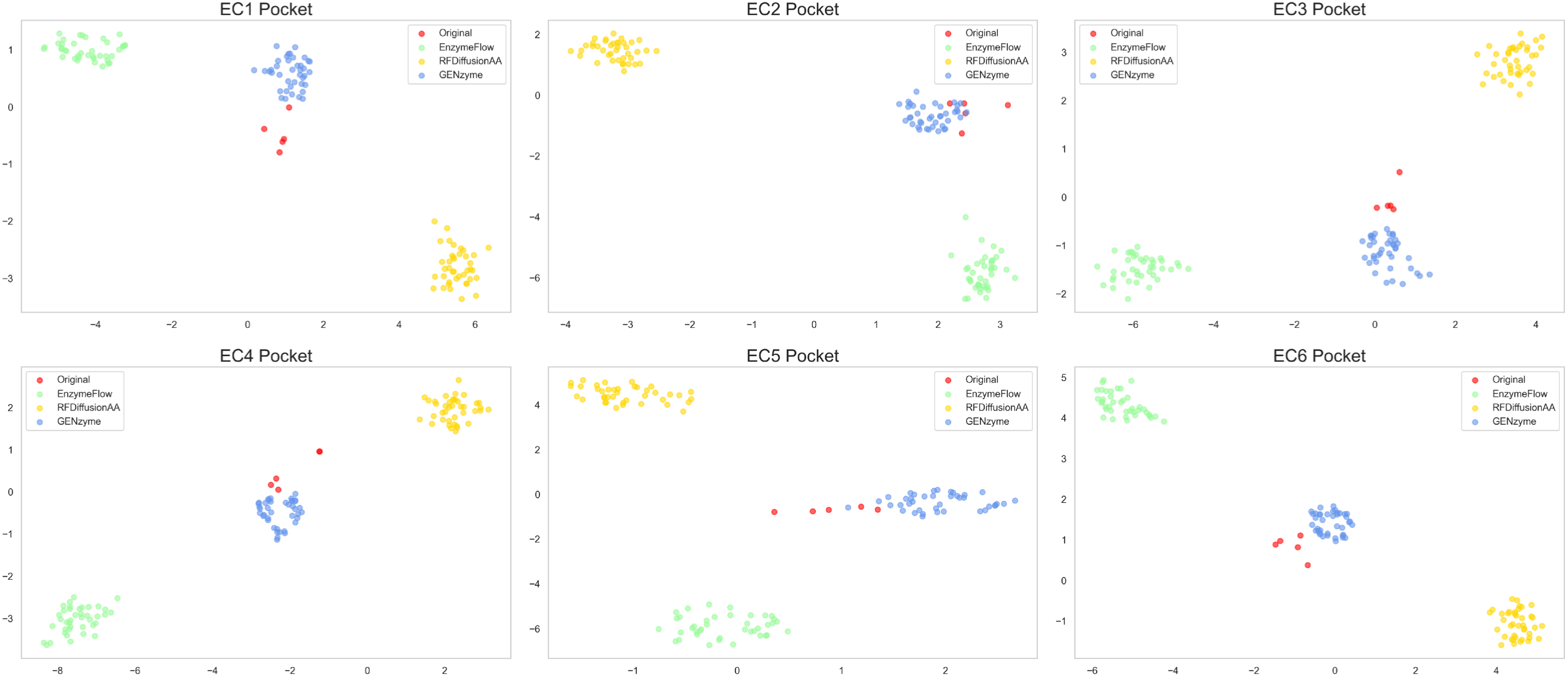}}
{
\vspace{-0.6cm}
  \caption{t-SNE visualization of \textbf{catalytic pocket embeddings} generated by ESM3. \textcolor{BrickRed}{Red denotes ground-truth catalytic pockets}, \textcolor{Goldenrod}{yellow denotes RFDiffusionAA-generated pockets}, \textcolor{Cerulean}{blue denotes \textsc{GENzyme}-generated pockets}, and \textcolor{LimeGreen}{green denotes EnzymeFlow-generated pockets}.}
  \label{fig:pocket_tsne}
}
\end{figure*}

\begin{figure*}[ht!]
\vspace{-0.2cm}
\centering
{
\includegraphics[width=1\textwidth]{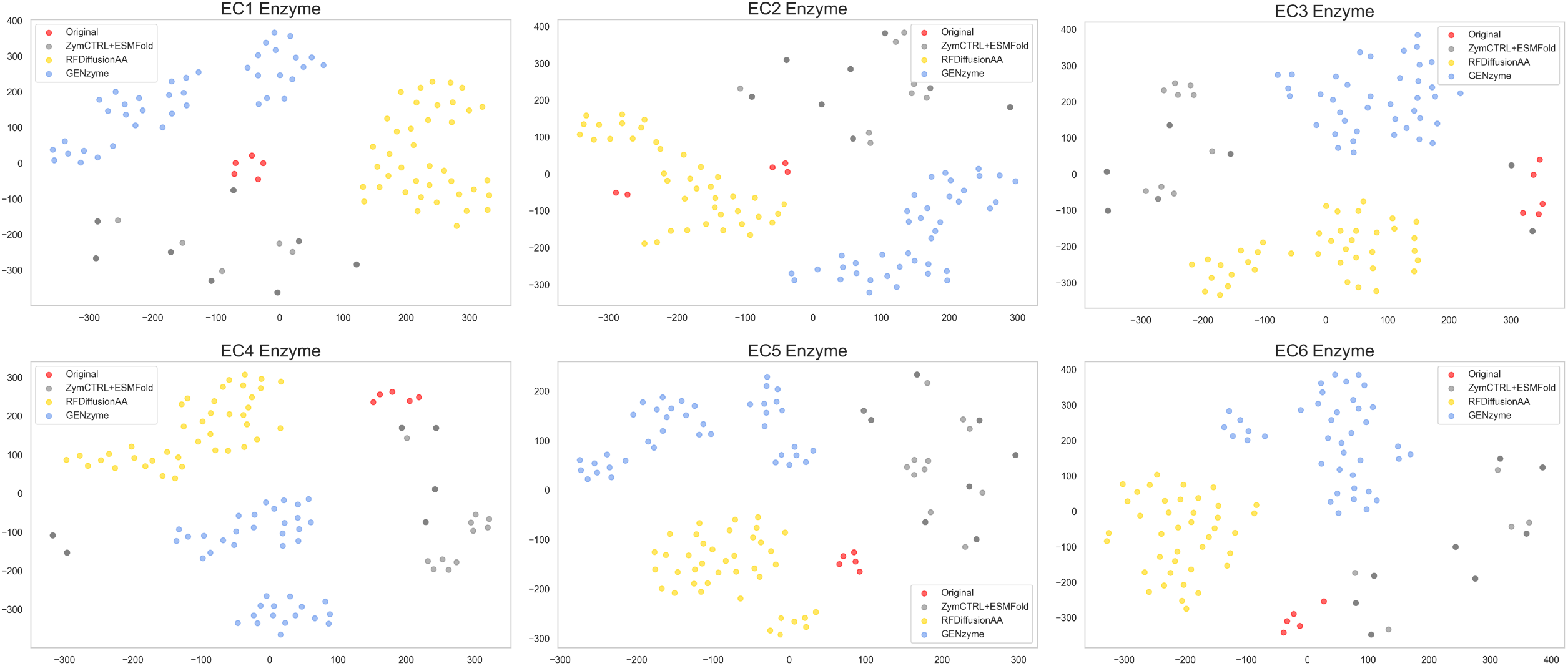}}
{
\vspace{-0.6cm}
  \caption{t-SNE visualization of \textbf{full enzyme embeddings} generated by ESM3. \textcolor{BrickRed}{Red denotes ground-truth catalytic enzymes}, \textcolor{Goldenrod}{yellow denotes RFDiffusionAA-generated enzymes}, \textcolor{Cerulean}{blue denotes \textsc{GENzyme}-generated enzymes}, and \textcolor{gray}{gray denotes ZymCTRL+ESMFold-generated enzymes}.}
  \label{fig:protein_tsne}
  \vspace{-0.3cm}
}
\end{figure*}

\vspace{-0.1cm}
\label{sec:tsne.embeddings}
Additionally, we evaluate the generated catalytic pockets and full enzymes by analyzing their sequence-structure embeddings in comparison to ground-truth ones. We use ESM3 \citep{hayes2024simulating} to co-encode both protein sequences and structures, though alternative methods like FoldSeek \citep{van2022foldseek} are also viable for this purpose. Fig.~\ref{fig:pocket_tsne} illustrates the tSNE visualization of catalytic pocket embeddings, and Fig.~\ref{fig:protein_tsne} presents the embeddings of full enzyme structures.

By analyzing Fig.~\ref{fig:pocket_tsne}, we observe that the \textsc{GENzyme}-generated catalytic pockets cluster closely with the ground-truth pockets, indicating high catalytic potential, as these active regions and catalytic sites are where reactions occur. This alignment suggests that \textsc{GENzyme} effectively maintains key catalytic features. In contrast, Fig.~\ref{fig:protein_tsne} shows that the inpainted enzymes cluster separately from the ground-truth enzymes, which could imply enhanced overall enzyme stability and functionality in the \textit{de novo} designs. Together, Fig.~\ref{fig:pocket_tsne} and Fig.~\ref{fig:protein_tsne} suggest that \textsc{GENzyme} preserves essential catalytic properties while potentially improving overall enzyme stability and functional diversity.

\vspace{-0.1cm}
\subsection{Enzyme Function Evaluation}
\vspace{-0.1cm}
A key question in enzyme design is how to \textit{quantitatively} assess whether the generated catalytic pockets and enzymes can effectively catalyze a chemical reaction. To address this, we evaluate enzyme functions by analyzing the optimal pH, enzyme kinetics, and mutation effects of the generated pockets and full enzymes—three metrics that can provide insight into their catalytic potential.


First, we evaluate the \textbf{optimal pH (\(\texttt{pH}_\texttt{opt}\))}, which refers to the specific pH level at which an enzyme achieves its highest catalytic activity. Although most enzymes perform optimally within a pH range of $6.0$ to $8.0$, some enzymes are specialized to function in highly acidic (\(\texttt{pH}_\texttt{opt}<5.0\)) or highly alkaline (\(\texttt{pH}_\texttt{opt}>9.0\)) conditions \citep{bisswanger2014enzyme}. The pH affects enzyme activity by altering the enzyme shape and the charges at both the active site and substrate. For therapeutic applications, it is important that the enzyme functions well at physiological pH levels, or within the specific pH range of the tissue or fluid where it will act \citep{langer2004designing, di2009serine, concolino2018enzyme}. Therefore, we first evaluate the \(\texttt{pH}_\texttt{opt}\) of the designed enzymes to assess their functional suitability. To predict the optimal pH \(\texttt{pH}_\texttt{opt}\), we use EpHod \citep{gado2023deep}.

In addition to optimal pH, enzyme kinetics is a critical parameter for evaluating enzyme function. Specifically, we assess the \textbf{turnover number (\(\texttt{k}_\texttt{cat}\))}, which describes the rate at which enzyme-catalyzed reactions occur. This metric provides insights into the enzyme’s catalytic mechanism, metabolic role, and responsiveness to potential inhibitors or activators \citep{wilkinson1961statistical, cornish2013fundamentals}. The \(\texttt{k}_\texttt{cat}\) value, or turnover number, indicates the number of substrate molecules converted to product per second at each active site, thus reflecting the enzyme’s catalytic speed and efficiency. 
To predict and evaluate these kinetic parameters for the generated enzymes, we employ UniKP \citep{yu2023unikp}, a tool that offers comprehensive insights into enzyme function and catalytic efficiency through kinetic profiling.

We also assess enzyme function by measuring the \textbf{rate of change of Gibbs free energy (\texttt{$\Delta\Delta G$}) under mutation}, which quantifies the impact of mutation effects on enzyme stability relative to the wild-type enzymes \citep{romero2009exploring, soskine2010mutational}. A negative \(\Delta\Delta G\) (\(\Delta\Delta G<0\)) indicates that the mutation increases stability, potentially enhancing enzyme fitness. A positive \(\Delta\Delta G\) (\(\Delta\Delta G>0\)) suggests decreased stability via mutation, which can reduce enzyme fitness. A zero \(\Delta\Delta G\) (\(\Delta\Delta G\approx0\)) implies that the mutation has little to no effect on stability. The enzyme fitness landscape, which maps the mutation effects on function and stability, can be better understood by analyzing \(\Delta\Delta G\), as it helps predict whether a mutation will enhance or degrade the enzyme catalytic performance \citep{notin2024proteingym}. To predict mutation effect \(\Delta\Delta G\), we use Mutate-Everything \citep{ouyang2024predicting}.

\begin{figure*}[ht!]
\vspace{-0.2cm}
\centering
{
\includegraphics[width=1\textwidth]{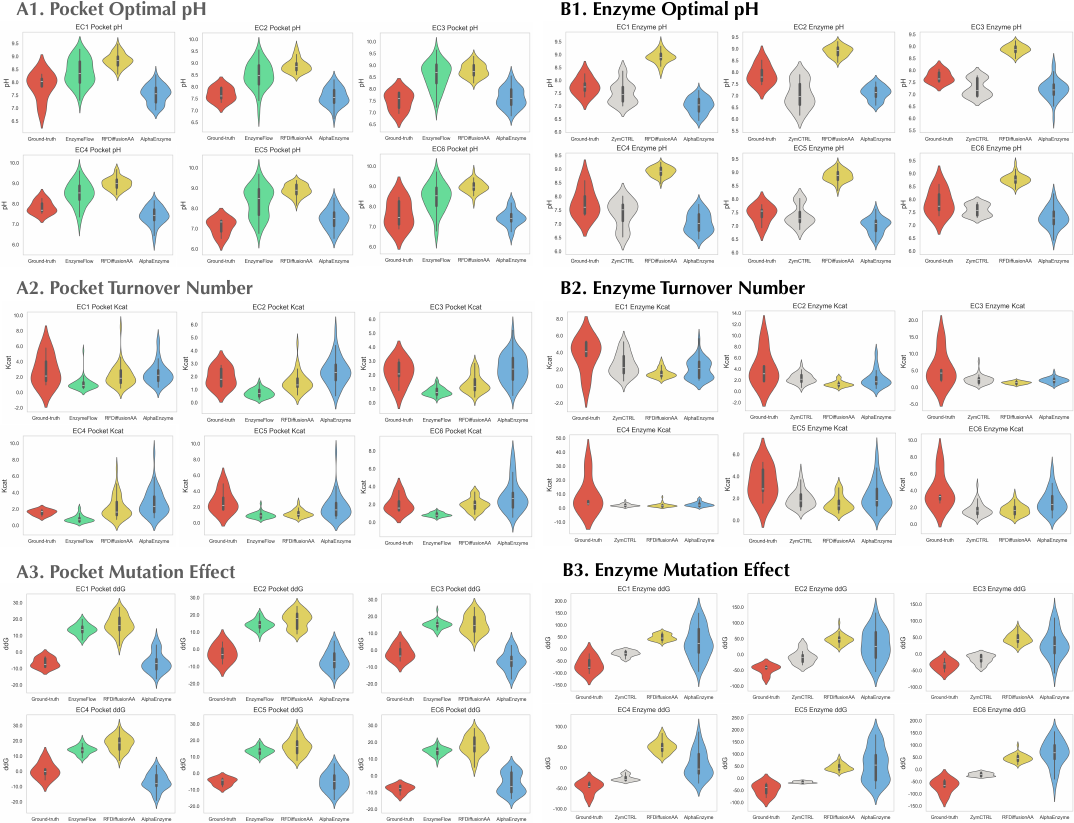}}
{
\vspace{-0.5cm}
  \caption{\textbf{Left: Pocket Function Evaluation}. (A1) Pocket optimal pH ($\texttt{pH}_\text{opt}$). (A2) Pocket turnover number (\(\texttt{k}_\texttt{cat}\)). (A3) Pocket mutation effects (\texttt{$\Delta\Delta G$}). \textbf{Right: Enzyme Function Evaluation}. (B1) Enzyme optimal pH ($\texttt{pH}_\text{opt}$). (B2) Enzyme turnover number (\(\texttt{k}_\texttt{cat}\)). (B3) Enzyme mutation effects (\texttt{$\Delta\Delta G$}). \textcolor{BrickRed}{Red denotes ground-truth}, \textcolor{LimeGreen}{green denotes EnzymeFlow-generated}, \textcolor{Goldenrod}{yellow denotes RFDiffusionAA-generated}, \textcolor{Gray}{gray denotes ZymCTRL-generated}, \textcolor{Cerulean}{blue denotes \textsc{GENzyme}-generated}.}
  \label{fig:pocket.enzyme.function}
  \vspace{-0.2cm}
}
\end{figure*}

In parallel to structural analysis, we assess the functional performance of \textsc{GENzyme}- and baseline-generated catalytic pockets in Fig.~\ref{fig:pocket.enzyme.function}(left), as well as inpainted and generated enzymes in Fig.~\ref{fig:pocket.enzyme.function}(right), examining functions at both local (catalytic regions) and global (entire enzyme) scales.

In evaluating optimal pH (\(\texttt{pH}_\texttt{opt}\)) at both the catalytic pocket and full-enzyme levels, we observe that \textsc{GENzyme} produces pockets and enzymes that align more closely with the optimal pH values of ground-truth examples compared to those generated by baselines. This alignment is important, as catalytic reactions typically occur most efficiently at an enzyme’s optimal pH. 

Regarding pocket turnover number (\(\texttt{k}_\texttt{cat}\)), \textsc{GENzyme} generates catalytic pockets and enzymes with improved turnover rates relative to the ground truth and baseline models at the pocket level and outperforms baselines at the enzyme level. This improvement suggests higher catalytic efficiency from \textsc{GENzyme}-generated catalytic pockets. 

For mutation effects (\(\Delta\Delta G\)), at the pocket level, \textsc{GENzyme}-generated pockets demonstrate stability under mutation, indicating that mutations have minimal impact on structural stability and thus are unlikely to hinder catalytic performance. Additionally, \textsc{GENzyme} shows greater resilience to mutation effects at the enzyme level, suggesting an overall enhancement in stability. This reduced sensitivity to mutations implies that \textsc{GENzyme} designs can yield enzymes with more consistent functional stability across various conditions, further supporting stable catalytic activity.


\vspace{-0.1cm}
\section{GENzyme Method}
\vspace{-0.1cm}
We introduce \textsc{GENzyme} approach in App.~\ref{app.preliminaries}, \ref{app.alphaenzyme.method}.

\vspace{-0.1cm}
\section{GENzyme Discussion}
\vspace{-0.1cm}

\textsc{GENzyme} envisions a comprehensive approach to generating enzymes for specific catalytic reactions, aiming to contribute to therapeutic solutions, decompose harmful substances, and enhance our understanding of metabolic pathways. While it represents an important step forward, \textsc{GENzyme} is not yet a complete solution and does not reach the level of models like RFDiffusionAA \citep{krishna2024generalized}, AlphaFold \citep{jumper2021highly, abramson2024accurate}, or AlphaProteo \citep{zambaldi2024novo}. However, we hope that \textsc{GENzyme} can serve as a foundation for future advancements, inspiring further research and progress in \textit{de novo} enzyme design, eventually leading to successful computational enzyme discovery and influencing therapeutics.

\vspace{-0.1cm}
\subsection{Model Generalizability}
\vspace{-0.1cm}
We are currently testing \textsc{GENzyme} on the latest 2024 enzyme and reaction datasets collected from Rhea \citep{bansal2022rhea}, MetaCyc \citep{caspi2020metacyc}, and Brenda \citep{schomburg2002brenda}. This will help us evaluate \textsc{GENzyme}'s performance and adaptability on newly collected, unseen data.

\vspace{-0.1cm}
\subsection{Limitations of Structure-Based Models for Catalytic Reaction Modeling}
\vspace{-0.1cm}
This section explores instances where structure-based protein and complex design models, including recent powerful methods like Chai-1 and AlphaFold3 \citep{Chai-1-Technical-Report, abramson2024accurate}, occasionally fall short in accurately modeling catalytic reactions. A notable limitation in these complex generation models, particularly for our \textit{de novo} enzyme design approach, is the requirement for an input protein sequence to generate the enzyme-substrate complex. However, for this analysis, we assume the enzyme sequence is known and focus solely on reaction modeling.

In our experiments using Chai-1 to model catalytic reactions, we observed a generally high success rate, where catalytic regions (active sites or catalytic pockets) remained stable and mostly unchanged throughout the catalytic process. Nonetheless, there are cases where Chai-1 does not maintain catalytic consistency, leading to significant alterations in catalytic regions.

\begin{figure*}[ht!]
\centering
{
\includegraphics[width=1\textwidth]{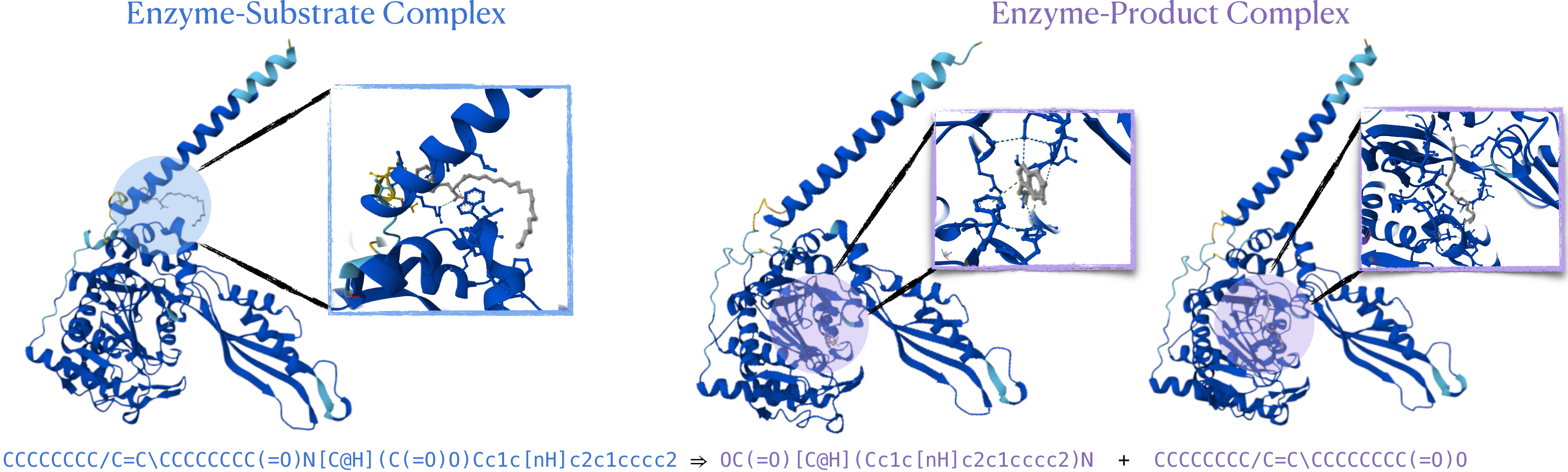}}
{
\vspace{-0.5cm}
  \caption{Catalytic reaction modeling for the enzyme of UniProt Q08BB2 using Chai-1.}
  \label{fig:chai1}
  \vspace{-0.2cm}
}
\end{figure*}

Fig.~\ref{fig:chai1} presents an example of such a failure, where Chai-1 does not preserve the original catalytic sites. The catalytic regions before the reaction (\textcolor{blue}{in blue}) shift considerably after the reaction (\textcolor{purple}{in purple}) for the resulting products. This illustrates that even advanced models like Chai-1 occasionally struggle with modeling catalytic reactions effectively. This finding highlights the need for a shift towards function-driven enzyme design, emphasizing catalytic activity as a primary objective—whether defined by specific reactions, as in our approach, or by enzyme classification systems like ZymCTRL’s EC-number-controlled generative strategy \citep{munsamy2022zymctrl}.

\vspace{-0.1cm}
\subsection{What Does GENzyme Aim to Address?}
\vspace{-0.1cm}
\textbf{Advancing \textit{De Novo} Enzyme Design for Unseen Reactions.} \textsc{GENzyme} seeks to address the challenge of designing enzymes for novel, unseen catalytic reactions. This could deepen our understanding of metabolic pathways and offer therapeutic benefits, including disease intervention. By combining \textit{de novo} catalytic pocket design, pocket inpainting, and protein evolutionary techniques, \textsc{GENzyme} attempts to generate new catalytic pockets both through computational design and biological mutation perspectives. This dual approach could make the algorithm more robust and generalizable to reactions that have not yet been experimentally observed.

\textbf{Shifting to Function-based Protein Design.} Unlike structure-based models such as RFDiffusionAA \citep{krishna2024generalized}, AlphaFold3 \citep{abramson2024accurate}, or Chai1 \citep{Chai-1-Technical-Report}, which prioritize generating proteins based on structures, \textsc{GENzyme} aligns with a function-driven strategy. The input in \textsc{GENzyme} is defined by the enzyme function—the catalytic reaction it is meant to perform—rather than merely its structure. This approach emphasizes not only binding performance, \textit{i.e.,} optimizing the enzyme-substrate binding affinities, but also ensures that the enzyme structure can facilitate the dynamic kinetic changes needed for catalysis. Although \textsc{GENzyme} encodes the protein function in a simplified way (via tokenized catalytic reactions), it introduces a novel direction for function-based protein design, where the function defines the structure.

\textbf{Addressing Data Scarcity in Enzyme Design.} A long-standing issue in enzyme design has been the lack of enzyme-reaction data with detailed structures or binding poses of enzyme-substrate complexes. Without this structural information, understanding how enzymes interact with substrates in the active site, and subsequently achieve catalysis, can been difficult. \textsc{GENzyme} attempts to address this challenge by generating synthetic enzyme-substrate complexes, which could provide insights into the mechanisms of catalysis and how enzymes facilitate specific reactions, helping bridge the gap in our understanding of enzyme-substrate interactions.

\textbf{Fine-tuned EnzymeESM for Representation Learning.} Beyond its focus on \textit{de novo} enzyme design for therapeutic applications, \textsc{GENzyme} also contributes to enzyme representation learning. Current deep learning methods often employ models like ESM2 \citep{lin2022language} or ProtT5 \citep{elnaggar2021prottrans} to embed enzymes for downstream tasks such as enzyme function prediction. Although these protein language models are pre-trained on millions of sequences, fine-tuning them for enzyme-specific tasks is necessary, similar to how large language models like LLaMa are fine-tuned for specific downstream applications \citep{howard2018universal, touvron2023llama}. In this context, we fine-tune ESM3 \citep{hayes2024simulating} on the EnzymeFill dataset, learning both enzyme sequences and structures. EnzymeESM is fine-tuned over 300,000 enzyme-reaction pairs, enabling tasks including enzyme catalytic pocket scaffolding, enzyme inverse folding, and enzyme embedding.

\vspace{-0.1cm}
\subsection{What is still missing in GENzyme?}
\vspace{-0.1cm}
\textbf{Virtual Screening in Computational Enzyme Discovery.} A key gap in \textsc{GENzyme}, and in the broader field of computational enzyme discovery, is the lack of robust virtual screening models. Recent works have attempted to address this using contrastive learning methods to model enzyme-reaction relationships both from sequence and structural perspectives \citep{mikhael2024clipzyme, yang2024care, hua2024reactzyme}. However, current screening models remain insufficiently robust. Traditional approaches, such as using tools like Vina to compute binding affinities between enzymes and substrate conformations \citep{trott2010autodock}, focus heavily on structural evaluations, which may not capture the full complexity of enzyme catalysis.
An ideal virtual screening model, in our view, should take an enzyme-reaction pair as input and output a confidence score ranging from 0 to 1, indicating how likely the enzyme is to catalyze the reaction. This concept has been hinted at in earlier work \cite{hua2024enzymeflow}. However, creating such a model presents significant challenges, particularly in generating meaningful negative enzyme-reaction pairs. For instance, CLIPZyme \citep{mikhael2024clipzyme} treats all non-positive enzyme-reaction pairs as negative examples, while ReactZyme \citep{hua2024reactzyme} creates negative pairs by mutating amino acids in positive data. These methods are inspired by CLIP models in computer vision, where a mismatched image-text pair can clearly be labeled as negative \citep{radford2021learning}. However, in enzyme-reaction prediction, an enzyme-reaction pair that does not exist in positive data could simply be an unobserved reaction, not necessarily a negative one. This limits the generalizability of current enzyme-reaction CLIP models, particularly for virtual screening of unseen enzyme-reaction pairs.
We believe that improving virtual screening models in computational enzyme discovery is more important than generative models alone, as robust screening will enable the identification of truly functional enzyme-reaction pairs, enhancing the accuracy of enzyme design efforts.

\textbf{Integration of AI-assisted Protein Evolution.}
Another missing aspect in \textsc{GENzyme} is the broader use of directed protein evolution, such as amino acid mutations, which are important for practical enzyme design \citep{arnold1998design, schmidt1999directed, eijsink2005directed, king2024computational}. Although \textsc{GENzyme} models enzyme-reaction co-evolution to account for dynamic changes in catalytic reactions, this is still insufficient from a practical, wet-lab perspective. In real-world experimental designs, researchers often start with enzyme mutations, rather than generating full enzyme structures from scratch.
A more practical approach would be to focus on motif scaffolding—replacing active regions of existing enzymes with newly generated catalytic pockets, while preserving the rest of the enzyme backbone structure \citep{hossack2023building}. This method would align more closely with wet-lab needs, where scaffold-based modifications are often preferred due to higher success rates in enzyme design experiments. By designing enzymes from a scaffolding perspective, \textsc{GENzyme} could increase the likelihood of successful wet-lab implementation, ultimately bridging the gap between computational predictions and experimental outcomes.

\vspace{-0.1cm}
\subsection{Broader Impact and Future Work}
\vspace{-0.1cm}
The \textit{de novo} enzyme design approach, while groundbreaking, naturally encounters limitations, particularly in verifying enzyme functionality. Although we are making strides toward addressing this through an enzyme evolution framework, our efforts remain in progress. Future work will focus on incorporating principles of enzyme engineering and site-specific mutations to enhance the functional reliability and catalytic efficiency of the \textit{de novo} enzymes generated. These improvements aim to build a more robust and validated pathway for enzyme design, bridging the gap between computational predictions and real-world enzymatic applications.

A key challenge in enzyme design lies in accurately modeling substrate-product transitions, which are central to understanding catalytic mechanisms. Current approaches primarily focus on static binding models, yet dynamic interactions during catalysis remain underexplored. Future work will involve developing methods to model these substrate-product transitions more effectively, enabling designs that capture the full catalytic process. This could significantly improve the precision of \textit{de novo} enzyme designs, bridging the gap between computational models and practical, functional enzymes capable of specific catalysis.

\newpage
\section*{Acknowledgement}
\vspace{-0.2cm}
Chenqing thanks to the FACS-Acuity Project of Canada (No.~10242) and CIFAR AI Chairs, Shuangjia thanks to the National Natural Science Foundation of China (No.~62402314) and Aureka Bio. We extend our gratitude to Bozitao Zhong, Zuobai Zhang, Peter Mikhael, Hannes Stark, Sitao Luan for their valuable discussions and insights. We would like to acknowledge Mila and Nvidia for providing computational resources for the protein and enzyme experiments.

\vspace{-0.2cm}
\section*{Correspondence}
\vspace{-0.2cm}
The paper corresponds to Chenqing Hua (\url{chenqing.hua@mail.mcgill.ca}), Doina Precup (\url{dprecup@cs.mcgill.ca}), and Shuangjia Zheng (\url{shuangjia.zheng@sjtu.edu.cn}).

\vspace{-0.2cm}
\section*{License}
\vspace{-0.2cm}
\textsc{GENzyme} is available under a Non-Commercial license.

\vspace{-0.2cm}
\section*{Contribution Statement} 
\vspace{-0.2cm}
Chenqing contributes to the overall project, including proposing the workflow, coding, conducting experiments, and writing. Jiarui is responsible for fine-tuning ESM3 on the EnzymeFill. Yong contributes the data collection for EnzymeFill. Odin contributes to experiment validation. Jian provides supervision for Jiarui. Rex contributes to providing computational resourses and writing. Wengong provides supervision. Guy and Doina provide support by securing computational resources, funding, and offering supervision. Shuangjia contributes to experimental validation, paper writing, and providing supervision.

\vspace{-0.2cm}
\section*{Competing Interests}
\vspace{-0.2cm}
The authors claim no competing interests.

\vspace{-0.2cm}
\section*{Reproducibility Statement}
\vspace{-0.2cm}
We provide our code and example scripts with demonstrations at \url{https://github.com/WillHua127/GENzyme}.

\vspace{-0.2cm}
\section*{Impact Statement}
\vspace{-0.2cm}
\textsc{GENzyme} aims to push the boundaries of \textit{de novo} enzyme design, with potential applications in therapeutic applications and disease management. The next works include: (1) improving enzyme screening and filtering methods, (2) incorporating protein engineering and mutation strategies into the model, (3) augmenting data with enzyme kinetics, (4) integrating molecular dynamics, such as bond breaking, to better represent catalytic reaction conditions, (5) validating results through wet-lab experiments, and (6) training large-scale models.

\vspace{-0.2cm}
\section*{Dataset Statement}
\vspace{-0.2cm}
We will soon release EnzymeFill data, in the spring or summer term of 2025.

\vspace{-0.2cm}
\section*{Collaboration Statement}
\vspace{-0.2cm}
We welcome collaborations, inquiries, and discussions. Please feel free to reach out to us at \url{chenqing.hua@mail.mcgill.ca} and \url{shuangjia.zheng@sjtu.edu.cn}. 
We are actively seeking experimentalists for wet-lab validation of \textit{de novo} enzyme designs, as well as institutes or companies interested in providing additional data or computational resources to support large-scale training of \textsc{GENzyme} model.

\newpage
\bibliography{iclr2025_conference}
\bibliographystyle{iclr2025_conference}

\newpage
\appendix

\section{Extended Related Work}
\subsection{Protein Evolution}
Protein evolution learns how proteins change over time through processes such as mutation, selection, and genetic drift \citep{pal2006integrated, bloom2009light}, which influence protein functions. Studies on protein evolution focus on understanding the molecular mechanisms driving changes in protein sequences and structures. \cite{zuckerkandl1965molecules} introduce the concept of the molecular clock, which postulates that proteins evolve at a relatively constant rate over time, providing a framework for estimating divergence times between species. \cite{depristo2005missense} show that evolutionary rates are influenced by functional constraints, with regions critical to protein function (\textit{e.g.}, active sites, binding interfaces) evolving more slowly due to purifying selection. This understanding leads to the development of methods for detecting functionally important residues based on evolutionary conservation.
Understanding protein evolution has practical applications in protein engineering. By studying how natural proteins evolve to acquire new functions, researchers design synthetic proteins with desired properties \citep{xia2004simulating, jackel2008protein}. Additionally, deep learning models increasingly integrate evolutionary principles to predict protein function and stability, design novel enzymes, and guide protein engineering \citep{yang2019machine, alquraishi2019end, jumper2021highly}.

\subsection{Protein Representation Learning}
Graph representation learning emerges as a potent strategy for learning about proteins and molecules, focusing on structured, non-Euclidean data \citep{satorras2021n, luan2020complete, luan2022revisiting, hua2022graph, hua2022high, luan2024graph, luan2024heterophilic}. 
In this context, proteins and molecules can be effectively modeled as 2D graphs or 3D point clouds, where nodes correspond to individual atoms or residues, and edges represent interactions between them \citep{gligorijevic2021structure, zhang2022protein, hua2023mudiff, zhang2024deep}. 
Indeed, representing proteins and molecules as graphs or point clouds offers a valuable approach for gaining insights into and learning the fundamental geometric and chemical mechanisms governing protein-ligand interactions.
This representation allows for a more comprehensive exploration of the intricate relationships and structural features within protein-ligand structures \citep{tubiana2022scannet, isert2023structure, zhang2024ecloudgen, yu2024fraggen}.

\subsection{Protein Function Annotation}
Protein function prediction aims to determine the biological role of a protein based on its sequence, structure, or other features. It is a crucial task in bioinformatics, often leveraging databases such as Gene Ontology (GO), Enzyme Commission (EC) numbers, and KEGG Orthology (KO) annotations \citep{bairoch2000enzyme, gene2004gene, mao2005automated}. Traditional methods like BLAST, PSI-BLAST, and eggNOG infer function by comparing sequence alignments and similarities \citep{altschul1990basic, altschul1997gapped, huerta2019eggnog}. Recently, deep learning has introduced more advanced approaches for protein function prediction \citep{ryu2019deep, kulmanov2020deepgoplus, bonetta2020machine}.
There are two major types of function prediction models, one uses only protein sequence as their input, while the other also uses experimentally-determined or predicted protein structure as input. Typically, these methods predict EC or GO annotations to approximate protein functions, rather than describing the exact catalyzed reaction, which is a limitation of these approaches.

\subsection{Generative Models for Protein and Pocket Design}
Recent advancements in generative models have advanced the field of protein design and binding pocket design, enabling the creation of proteins or binding pockets with desired properties and functions \citep{yim2023fast, yim2023se, chu2024all, hua2024effective, abramson2024accurate}. For example, RFDiffusion \citep{watson2023novo} employs denoising diffusion in conjunction with RoseTTAFold \citep{baek2021accurate} for \textit{de novo} protein structure design, achieving wet-lab-level generated structures that can be extended to binding pocket design. RFDiffusionAA \citep{krishna2024generalized} extends RFDiffusion for joint modeling of protein and ligand structures, generating ligand-binding proteins and further leveraging MPNNs for sequence design. Additionally, FAIR \citep{zhang2023full} and PocketGen \citep{zhang2024pocketgen} use a two-stage coarse-to-fine refinement approach to co-design pocket structures and sequences. Recent models leveraging flow matching frameworks have shown promising results in these tasks. For instance, FoldFlow \citep{bose2023se} introduces a series of flow models for protein backbone design, improving training stability and efficiency. FrameFlow \citep{yim2023fast} further enhances sampling efficiency and demonstrates success in motif-scaffolding tasks using flow matching, while MultiFlow \citep{campbell2024generative} advances to structure and sequence co-design. These flow models, initially applied to protein backbones, have been further generalized to binding pockets. For example, PocketFlow \citep{zhang2024generalized} combines flow matching with physical priors to explicitly learn protein-ligand interactions in binding pocket design, achieving stronger results compared to RFDiffusionAA.
And EnzymeFlow \citep{hua2024enzymeflow} introduces a flow-based generative model, leveraging enzyme-reaction co-evolution and structure-based pre-training for enzyme catalytic pocket generation.

\textbf{\textsc{GENzyme} Contributions.} The first major contribution of \textsc{GENzyme} is its ability to perform \textit{de novo} enzyme design, generating catalytic pocket structures and full enzyme structures capable of catalyzing previously unseen reactions. The second contribution is \textsc{GENzyme} generation of synthetic enzyme-substrate binding complex data, which aids in a deeper understanding of enzyme-substrate interactions and metabolic processes. The third contribution is the fine-tuned protein language models for enzyme representation learning, optimizing them for enzyme-specific tasks such as catalytic pocket inpainting, enzyme inverse folding, and enzyme representation learning.

\section{Preliminaries}
\label{app.preliminaries}
\subsection{Enzyme-Reaction Notations}
Following \cite{jumper2021highly, yim2023fast}, we refer to the enzyme structure as the backbone atomic coordinates of each residue. An enzyme with number of residues \(N_E\) can be parameterized into SE(3) residue frames \(\{(x_i, r_i, a_i)\}_{i=1}^{N_E}\), where \(x_i \in \mathbb{R}^3\) represents the position (translation) of the \(C_\alpha\) atom of the \(i\)-th residue, \(r_i \in \text{SO(3)}\) is a rotation matrix defining the local frame relative to a global reference frame, and \(a_i \in \{1, \dots, 20\}\)  denotes the amino acid type. We refer to the residue block as \(E_i = (x_i, r_i, a_i)\), and the entire enzyme is described by a set of residues \(\mathbf{E} = \{E_i\}_{i=1}^{N_E}\) . 

A chemical reaction, \small $m_r$: $m_1 \xRightarrow{\mathbf{E}} m_2$, \normalsize describes a process in which a substrate molecule \(m_1\) is transformed into a product molecule \(m_2\), catalyzed by the enzyme \(\mathbf{E}\). The reaction, substrate, and product can be represented by canonical SMILES. The catalytic pocket \(\mathbf{E}^P = \{E^P_i\}_{i=1}^{N_P}\), consisting of \(N_P\) residues, is the active site within the enzyme \(\mathbf{E}\) where the substrate \(m_1\) binds, forming an enzyme-substrate complex \(\mathbf{C} = [\mathbf{E}, m_1]\), facilitating the reaction. Additionally, the enzyme scaffold \(\mathbf{E}^S = \{E^S_i\}_{i=1}^{N_S}\), consisting of \(N_S\) residues, complements the catalytic pocket, such that \(\mathbf{E} = \mathbf{E}^P \cup \mathbf{E}^S\) where \(N_E = N_P + N_S\).

\subsection{Conditional Flow Matching}
Flow matching describes a process where a flow transforms a simple distribution \( p_0 \) into the target data distribution \( p_1 \) \citep{lipman2022flow}. The goal in flow matching is to train a neural network \( v_\theta(\epsilon_t, t) \) that approximates the vector field \( u_t(\epsilon) \), which measures the transformation of the distribution \( p_t(\epsilon_t) \) as it evolves toward \( p_1(\epsilon_t) \) over time \( t \in [0, 1) \). The process is optimized using a regression loss defined as \( \mathcal{L}_{\text{FM}} = \mathbb{E}_{t\sim \mathcal{U}[0,1], p_t(\epsilon_t)} \|v_\theta(\epsilon_t, t) - u_t(\epsilon)\|^2 \).
However, directly computing \( u_t(\epsilon) \) is often intractable in practice. Instead, a conditional vector field \( u_t(\epsilon | \epsilon_1) \) is defined, and the conditional flow matching objective is computed as \( \mathcal{L}_{\text{CFM}} = \mathbb{E}_{t\sim \mathcal{U}[0,1], p_t(\epsilon_t)} \|v_\theta(\epsilon_t, t) - u_t(\epsilon | \epsilon_1)\|^2 \). Notably, \( \nabla_\theta \mathcal{L}_\text{FM} = \nabla_\theta \mathcal{L}_\text{CFM} \).

During inference or sampling, an ODEsolver, \textit{e.g.}, Euler method, is typically used to solve the ODE governing the flow, expressed as \( \epsilon_1 = \texttt{ODEsolver}(\epsilon_0, v_\theta, 0, 1) \), where \( \epsilon_0 \) is the initial data and \( \epsilon_1 \) is the generated data.
In actual training, rather than directly predicting the vector fields, it is more common to use the neural network to predict the final state at \( t=1 \), then interpolates to calculate the vector fields. This approach has been shown to be more efficient and effective for network optimization \citep{yim2023fast, bose2023se, campbell2024generative}.

\subsubsection{Continuous Variable Trajectory}
Given the predictions for translation \(\hat{x}_1\) and rotation \(\hat{r}_1\) at \(t=1\), we can interpolate and their corresponding vector fields are computed as follows:
\begin{equation}
    {v}_\theta(x_t, t) = \frac{\hat{x}_1 - x_t}{1-t}, \quad {v}_\theta(r_t, t) = \frac{\log_{r_t} \hat{r}_1}{1-t}.
\end{equation}
The sampling or trajectory can then be computed using Euler steps with a step size \(\Delta t\), as follows:
\begin{equation}
    x_{t + \Delta t} = x_t + {v}_\theta(x_t, t) \cdot \Delta t, \quad r_{t + \Delta t} = r_t + {v}_\theta(r_t, t) \cdot \Delta t,
\end{equation}
where the prior of $x_0, r_0$ are chosen as the uniform distribution on $\mathbb{R}^3$ and SO(3), respectively.

\subsubsection{Discrete Variable Trajectory}
For the discrete variables, we use continuous time Markov chains (CTMC). 

\textbf{Continuous Time Markov Chain.}
A sequence trajectory $\epsilon_t$ over time $t \in [0, 1]$ that follows a CTMC alternates between resting in its current state and
periodically jumping to another randomly chosen state. The frequency and destination of the jumps are determined by the rate matrix $R_t \in \mathbb{R}^{N \times N}$ with the constraint its off-diagonal elements are non-negative. The probability of $\epsilon_t$ jumping to a different state $s$ follows $R_t(\epsilon_t, s)\mathrm{d}t$ for the next infinitesimal time step $\mathrm{d}t$. We can express the transition probability as
\begin{equation}
    p_{t + \mathrm{d}t}(s | \epsilon_t) = \delta \{\epsilon_t, s\} + R_t(\epsilon_t, s)\mathrm{d}t,
\end{equation}
where \(\delta(\text{a}, \text{b})\) is the Kronecker delta, equal to \(1\) if \(\text{a} = \text{b}\) and \(0\) if \(\text{a} \neq \text{b}\), and $R_t(\epsilon_t, \epsilon_t) = -\sum_{\gamma \neq \epsilon} (\epsilon_t, \gamma)$ \citep{campbell2024generative}. Therefore, $p_{t+\mathrm{d}t}$ is a Categorical distribution with probabilities \(\delta(\epsilon_t, \cdot) + R_t(\epsilon_t, \cdot)\mathrm{d}t\) with notation $s \sim \text{Cat}(\delta(\epsilon_t, s) + R_t(\epsilon_t, s)\mathrm{d}t)$.

For finite time intervals $\Delta t$, a sequence trajectory can be
simulated with Euler steps following:
\begin{equation}
    \epsilon_{t+\Delta t} \sim \text{Cat}(\delta(\epsilon_t, \epsilon_{t+\Delta t}) + R_t(\epsilon_t, \epsilon_{t+\Delta t})\Delta t).
\end{equation}
The rate matrix $R_t$ along with an initial distribution $p_0$ define CTMC.
Furthermore, the probability flow $p_t$ is the marginal distribution of $\epsilon_t$ at every time $t$, and we say the rate matrix $R_t$ generates $p_t$ if $\partial_t p_t = R^T_tp_t, \forall t \in [0,1]$.

In the actual training, \cite{campbell2024generative} show that we can train a neural network to approximate the true denoising distribution using the standard cross-entropy:
\begin{equation}
    \mathcal{L}_\text{CE} = \mathbb{E}_{t\sim \mathcal{U}[0,1], p_t(\epsilon_t)} [\log p_\theta(\epsilon_1 | \epsilon_t)].
\end{equation}

\textbf{Rate Matrix for Inference.}
The conditional rate matrix $R_t(\epsilon_t, s|s_1)$ generates the conditional flow $p_t(\epsilon_t | \epsilon_1)$. 
And $R_t(\epsilon_t, s) = \mathbb{E}_{p_1(\epsilon_1 | \epsilon_t)}[R_t(\epsilon_t, s | \epsilon_1)]$, for which the expectation is taken over $p_1(\epsilon_1|\epsilon_t) = \frac{p_t(\epsilon_t|\epsilon_1)p_1(\epsilon_1)}{p_t(\epsilon_t)}$. 
With the conditional rate matrix, the sampling can be performed:
\begin{equation}
\begin{split}
    & R_t(\epsilon_t, \cdot) \leftarrow \mathbb{E}_{p_1(\epsilon_1|\epsilon_t)}[R_t(\epsilon_t, \cdot | \epsilon_1)], \\
    & \epsilon_{t+\Delta t} \sim \text{Cat}(\delta(\epsilon_t, \epsilon_{t+\Delta t}) + R_t(\epsilon_t, \epsilon_{t+\Delta t})\Delta t).
\end{split}
\end{equation}
The rate matrix generates the probability flow for discrete variables.

\cite{campbell2024generative} define the conditional rate matrix starting with
\begin{equation}
    R_t(\epsilon_t, s | \epsilon_t) = \frac{\text{ReLU}(\partial_t p_t(s | \epsilon_1)- \partial_t p_t(\epsilon_t | \epsilon_1))}{N \cdot p_t(\epsilon_t | \epsilon_1)}.
\end{equation}
In practice, the closed-form of conditional rate matrix with \textit{masking state} \mask \ is defined as:
\begin{equation}
    R_t(\epsilon_t, s| \epsilon_1) = \frac{\delta(\epsilon_1, s)}{1-t} \delta(\epsilon_t, \text{ \mask}).
\end{equation}

\subsection{Discrete Diffusion Probabilistic Modeling}
The discrete diffusion models \citep{austin2021structured, lou2023discrete, sun2022score, campbell2022continuous,zheng2023reparameterized} can be generally defined by a sequential process of progressive noisy variables $z_t \in V$ from the categorical variable $z_0 \in V$. Denote the one-hot (row) vector of $z_t$ as $\vz_t \in \{0,1\}^{|V|}$,  in the discrete-time case \citep{austin2021structured}, the forward marginal probability of $\vz_t$ at time $t$ has the following form as a composition of Markov kernel defined by $\mQ_t ~(t=1,2\dots, T)$:
\begin{equation}\label{eq:dd-forward}
    q(\vz_t| \vz_0) = \cat\left(\vz_t; \vz_0\bar{\mQ}_t \right) \triangleq \cat\left(\vz_t; \vz_0\mQ_1\cdot \dots\cdot \mQ_t  \right),
\end{equation}
where $\mQ_t$ indicates the transition probability matrix for time $t$ represented by $[\mQ_t]_{ij} = q(z_t=j|z_{t-1}=i)$, and $\cat(\cdot; \vp), \vp \in \Delta^{|V|}$ indicates the categorical distribution with probability and $ \Delta^{|V|}$ is the $|V|$-simplex. Eq.~\ref{eq:dd-forward} also induce the form of the marginal distribution for $\forall t > s$ is $q(\vz_t| \vz_s) = \cat\left(\vz_t; \vz_s \bar{\mQ}_{t|s} \right) \triangleq \cat\left(\vz_t; \vz_s \mQ_{s+1}\cdot \dots\cdot \mQ_{t} \right)$.  Correspondingly, the posterior $q(\vz_{s} | \vz_t, \vz_0)$ can be obtained by the reverse process \citep{austin2021structured}:
\begin{equation}\label{eq:dd-backward}
    q(\vz_{s} | \vz_t, \vz_0) = \frac{q(\vz_{t} | \vz_{s}, \vz_0) q(\vz_{s} | \vz_0)}{q(\vz_{t} | \vz_0) }=
    \cat \left(\vz_{s};
        \frac{\vz_t{\mQ}_{t|s}^\top   \odot \vz_0\bar{\mQ}_{s}}{\vz_0 \bar{\mQ}_t \vz_t^\top} \right), \forall~ s<t.
\end{equation}

Both \citet{zhao2024improving} and \citet{shi2024simplified} discuss how the discrete-time diffusion process can be generalized to the time domain $t \in [0, 1]$, akin to the diffusion over continuous space \citep{song2020score}, by demonstrating the continuous-time limit as $T \to \infty$. Notably, when the stationary distribution is explicitly specified (denoted as $\vp \in \Delta^{|V|}$), we can choose a \textit{state-independent} transition kernel in the simple form:
$
{\mQ}_{t|s} \triangleq \left[\alpha(s)^{-1}\alpha(t)\mI + (1-\alpha(s)^{-1}\alpha(t))\vone^\top \vp\right],
$
thus simplifying the continuous-time forward marginal to:
\begin{equation}\label{eq:dd-interp}
    q(\vz_{t} | \vz_s) = \cat\left(\vz_t; \frac{\alpha(t)}{\alpha(s)}\vz_s + (1-\frac{\alpha(t)}{\alpha(s)})\vp \right),  \forall~0\le s<t< 1,
\end{equation}
where $\alpha(t) \in [0,1)$ is a  strictly monotone decreasing function with $\alpha_0=1$ and $\alpha_1 \to 0$. The equation above demonstrates that the discrete diffusion, when defined with an explicit stationary distribution, can be viewed as an interpolation between two categorical distributions controlled by $\alpha(t)$.
According to Eq.~\ref{eq:dd-backward}, the reverse process of diffusion defined in Eq.~\ref{eq:dd-interp} takes the following form for the posterior distribution, where $0\le s<t< 1$: 
\begin{equation}\label{eq:dd-interp-rev}
q(\vz_s | \vz_t, \vz_0) = 
\cat\left( \vz_s ; \frac{ \left[\mu(t,s)\vz_t  + (1-\mu(t,s)) \lambda_{\vz_t}(\vp) \vone \right] \odot \left[\alpha(s)\vz_0 + (1-\alpha(s))\vp\right] }{ \alpha(t) \lambda_{\vz_t}(\vz_0) + (1-\alpha(t)) \lambda_{\vz_t}(\vp)} \right),
\end{equation}
where $\mu(t,s) \triangleq \alpha(s)^{-1}\alpha(t)>0$ and indicator function $\lambda_{\vz_t}(\cdot) \triangleq  \langle{\vz_t}, \cdot \rangle $ for concision.

\subsubsection{Conditional Masked Diffusion Language Model}
Masked diffusion \citep{austin2021structured, lou2023discrete, shi2024simplified, sahoo2024simple, lu2024structure} represents a special case in which the transition includes an ``absorbing state'', denoted as \mask. In this formulation, the stationary distribution in Eq.~\ref{eq:dd-forward} assigns all probability mass to the unique special token \mask, such that $ P (z = \mask) = 1 $ and $ P (z \neq \mask) = 0 $. For convenience, we define $\vp_M \in \{0, 1\}^{|\bar{V}|}$ ($\bar{V} \triangleq V \cup \{\mask\}$) as the one-hot vector representing \mask. In masked diffusion, the stochastic forward process maps  $z_0  \to \mask$ and remains in this state thereafter (i.e., ``absorbing''). Conversely, the reverse process gradually unmasks (denoises) the \mask token to produce the data sample $ \vz_0 $, where $s<t$:
\begin{equation}\label{eq:mask_back}
    q(\vz_s | \vz_t, \vz_0) = \cat\left(\vz_s; [\beta(s, t) + (1-\lambda_M(\vz_t))(1-\beta(s,t))]\vz_t + \lambda_M(\vz_t) (1-\beta(s,t))\vz_0 \right),
\end{equation}
where $\beta(s, t) = \frac{1-\alpha(s)}{1-\alpha(t)}$ and $\lambda_M(\vz_t) = \langle \vp_M, \vz_t\rangle$. Eq.~\ref{eq:mask_back} implies when $\vz_t \neq {\mask}$, the backward process simply copies the unmasked token by $\vz_s \gets \vz_t$, i.e. $q(\vz_s | \vz_t, \vz_0) =  \cat\left(\vz_s; \vz_t\right)$; otherwise the probability mass interpolates between $\vp_M$ and $\vz_0$. The posterior $q(\vz_s | \vz_t, \vz_0)$ can be approximated by $p_\theta(\vz_s | \vz_t)$ using re-parameterization: $p_\theta(\vz_s | \vz_t) = q(\vz_s | \vz_t, \vu_\theta(t, \vz_t))$, where the neural net $\vu_\theta \in \Delta^{|\bar{V}|}$ is a neural network that outputs a probability vector that remains in $\Delta^{|\bar{V}|}$.

For inpainting generation, we now consider the conditional case of masked diffusion. Given the amino acid types $\vc$, our goal is to sample structure tokens through Eq.~\ref{eq:mask_back}, utilizing a conditional posterior $q(\vz_s | \vz_t, \vz_0; \vc)$. This posterior can be re-parameterized similarly by incorporating the condition into the backbone model, resulting in $ p_\theta(\vz_s | \vz_t; \vc) = q(\vz_s | \vz_t, \vu_\theta(t, \vz_t, \vc)) $. 
To achieve this goal, the reverse process simulated by $p_\theta(\vz_s | \vz_t; \vc)$ must effectively approximate the data distribution $p(\vz|\vc)$. A feasible training objective is to optimize the estimation of the conditional ELBO within the continuous-time integral, resulting in the following loss:
\begin{equation}
    \label{eq:obj}
    \mathcal{L}(\theta) =  \E_{\vc, \vz_0}\left\{\int_{t\in[0,1)}  \E_{\vz_t \sim q(\vz_t|\vz_0)}
    \left[\frac{1}{1-\alpha(t)}\frac{\partial \alpha(t)}{\partial t}
        \lambda_M(\vz_t)  \log \langle \vu_\theta(t, \vz_t, \vc),\vz_0\rangle
    \right] \text{dt} \right\},
\end{equation}
where $\vz_0$ is sampled from the learned encoder $q_\phi(\vz|\vx)$ with the corresponding amino acid condition $\vc$ from the data distribution $p(\vx, \vc)$, and $\lambda_M(\vz_t)$ implies the loss is only applied for the latents $\forall t, \textrm{s.t. } \vz_t =\mask$. In practice, we can employ Monte Carlo estimation to compute the integral.

\subsubsection{Bidirectional Encoder as denoising network}
We now discuss the implementation of the conditional denoising network using bidirectional encoder language models, such as BERT \citep{devlin2018bert}. First, consider the sequential generalization of masked diffusion with a sequence of categorical variables. Let $\vz_t$ now be a sequence of discrete structure tokens $\left[\vz_{t,[1]}, \vz_{t, [2]}, \dots, \vz_{t, [L]}\right]$ where $\vz_{t, [i]} \in \bar{V}, \forall i=1,\dots,L$. Due to the interpolation scheme of Eq.~\ref{eq:dd-interp} and Eq.~\ref{eq:mask_back}, we assume conditional independence and factorize the posterior distribution $p_\theta(\vz_s | \vz_t; \vc)$ across the $L$ output tokens, such that $p_\theta(\vz_s | \vz_t, \vc) = \prod_{i=1}^L p_\theta(\vz_{s,[i]} | \vz_t, \vc)= \prod_{i=1}^L q\left(\vz_{s,[i]} | \vz_{t,[i]}, \vu_{\theta, [i]}(t, \vz_t, \vc)\right)$ where $\vu_{\theta, [i]}(t, \vz_t, \vc)$ represents the $i$-th output channel of neural network. 
This implement coincides with the BERT-style transformer architecture and allow us to take advantage of existing protein foundation model, for example ESM3 \citep{hayes2024simulating}. For a sequence of tokens, the masked $\log$-term in the training objective from Eq.~\ref{eq:obj} is replaced by the summation: $ \sum_{i=1}^L \lambda_M(\vz_{t, [i]})  \log \langle \vu_{\theta, [i]}(t, \vz_t, \vc),\vz_{0,[i]}\rangle$, with notations the same as defined above.

\textbf{Modifications.} 
The following are special considerations for the network:
(1) \textit{Position-coupled encoding.} Unlike general translation problem, \textsc{GENzyme} maintain strict position-to-position correspondence between amino acid types and the latent tokens\footnote{The underlying co-evolutionary relationships between residues are fully shared across both amino acid types and its spatial patterns. }
. This inductive bias enables us to construct the input embedding for all position $i$ as follows:
    $\ve_{[i]} = f_\theta[e_z(\vz_{t,[i]}) + e_c(\vc_{[i]}) + e_t(t)] \in \R^{D},
    $
where $e_z:|\bar V| \mapsto \R^{D}, e_c:|\gS| \mapsto \R^{D}, e_t: \R \mapsto \R^{D}$ are the embedding functions and $f_\theta$ is a linear transformation. (2) \textit{Copying}. The unmasked tokens $\vz_{t,[i]} \neq \mask$ remain the same in spite of the model output. (3) \textit{Zero-out} \mask. Since $\vu_\theta$ parameterize the approximated clean data $\vz_0$ (fully unmasked), the \mask \ token cannot present in the output and its probability should be zero-out. This is equivalent to adding $-\infty$ to the logit. In our study, the pre-trained LM head of ESM3 is replaced with a randomly initialized head with augmented vocabulary ($\bar{V}$) during fine-tuning.

\section{GENzyme Method}
\label{app.alphaenzyme.method}
\textsc{GENzyme} is an end-to-end, reaction-conditioned, three-stage enzyme design model trained on the EnzymeFill dataset. It includes a catalytic pocket structure generation and sequence design module, a pocket inpainting and enzyme inverse folding module, and a pocket-specific enzyme-substrate binding module. \textsc{GENzyme} generates enzymes conditioned on the SMILES representations of substrates and products, enabling function-driven enzyme design for specific catalytic reactions.

In this section, we discuss the workflow and key components of the \textsc{GENzyme} architecture, including: (1) EnzymeFill and \textsc{GENzyme} training data, (2) the catalytic pocket generation and sequence design module, (3) the catalytic pocket inpainting and enzyme inverse folding module, and (4) the substrate binding and screening module.

\subsection{GENzyme Training Data}
\begin{figure*}[ht!]
\vspace{-0.2cm}
\centering
{
\includegraphics[width=1\textwidth]{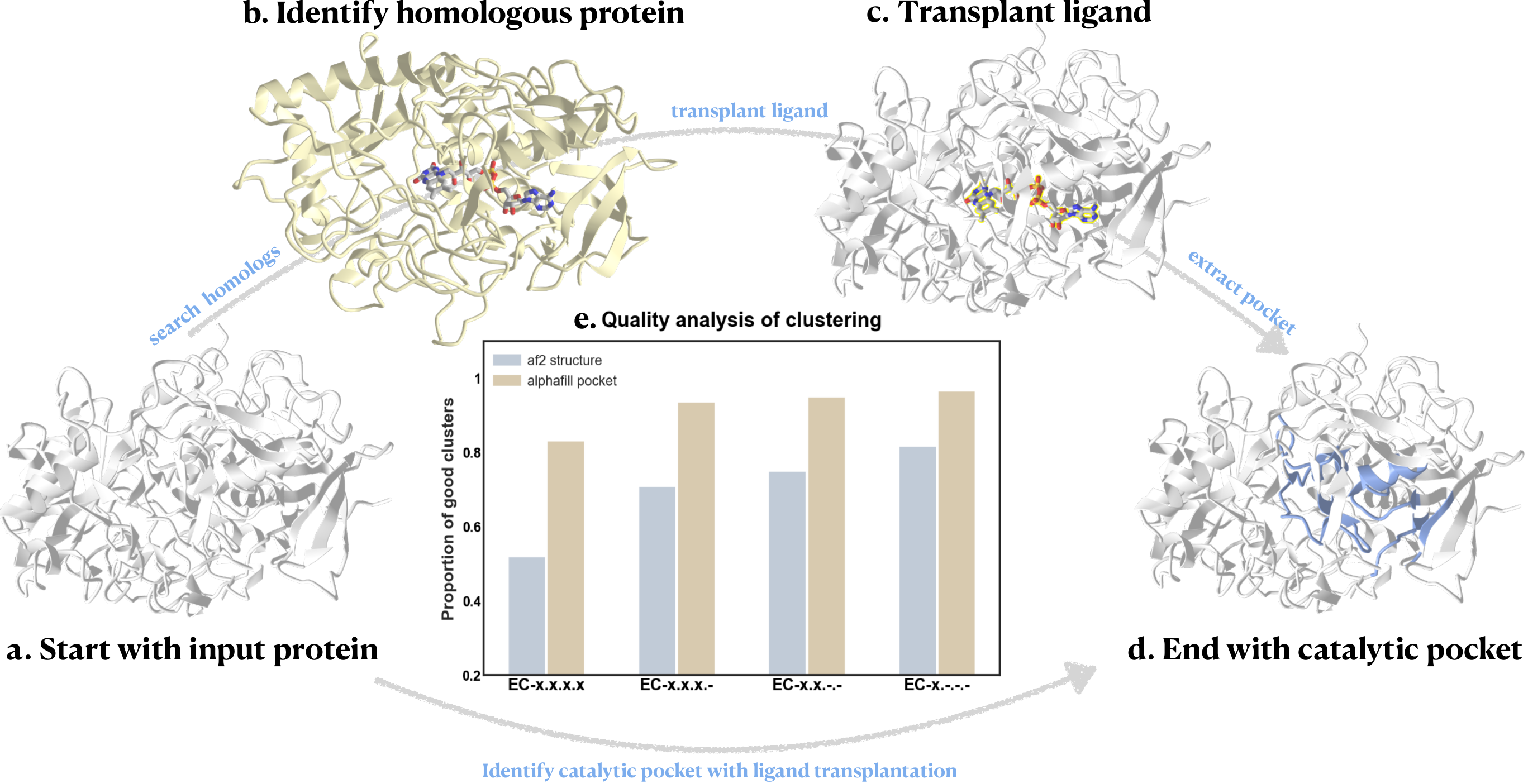}}
{
\vspace{-0.5cm}
  \caption{\textbf{EnzymeFill dataset construction with ligand transportation}. Starting with \textbf{(a)} an input enzyme structure from AlphaDB (\texttt{Uniprot P9WMV9}), the process \textbf{(b)} identifies a homologous protein in PDB-REDO (\texttt{PDB 1COY}), then \textbf{(c)} transplants ligands from the homologous protein complexes to the target enzyme. \textbf{(d)} A catalytic pocket is identified using a pre-defined radius of \small$10\mathring{\text{A}}$\normalsize. \textbf{(e)} Quality analysis of clustering between enzyme catalytic pockets and full structures, where strong clusters indicate high functional concentration.}
  \label{fig:pocket.extraction}
  \vspace{-0.5cm}
}
\end{figure*}

\textsc{GENzyme} is trained on EnzymeFill dataset. EnzymeFill is a curated and validated dataset of enzyme-reaction pairs with valid catalytic pocket structures \citep{hua2024enzymeflow}, comprising catalytic reactions collected from the Rhea \citep{bansal2022rhea}, MetaCyc \citep{caspi2020metacyc}, and Brenda \citep{schomburg2002brenda} databases. It contains a total of $328,192$ enzyme-reaction pairs, including $145,782$ unique enzymes and $17,868$ unique reactions.

Catalytic pockets in EnzymeFill are extracted from AlphaDB structures using AlphaFill and ligand transplantation \citep{hekkelman2023alphafill}, as demonstrated in Fig.~\ref{fig:pocket.extraction}. Fig.~\ref{fig:pocket.extraction}(e) highlights that enzyme catalytic pockets capture functional information more effectively than full enzyme structures, which supports the approach of focusing on catalytic pocket generation rather than full structure generation.

\begin{figure*}[htbp!]
\vspace{-0.3cm}
\centering
{
\includegraphics[width=0.6\textwidth]{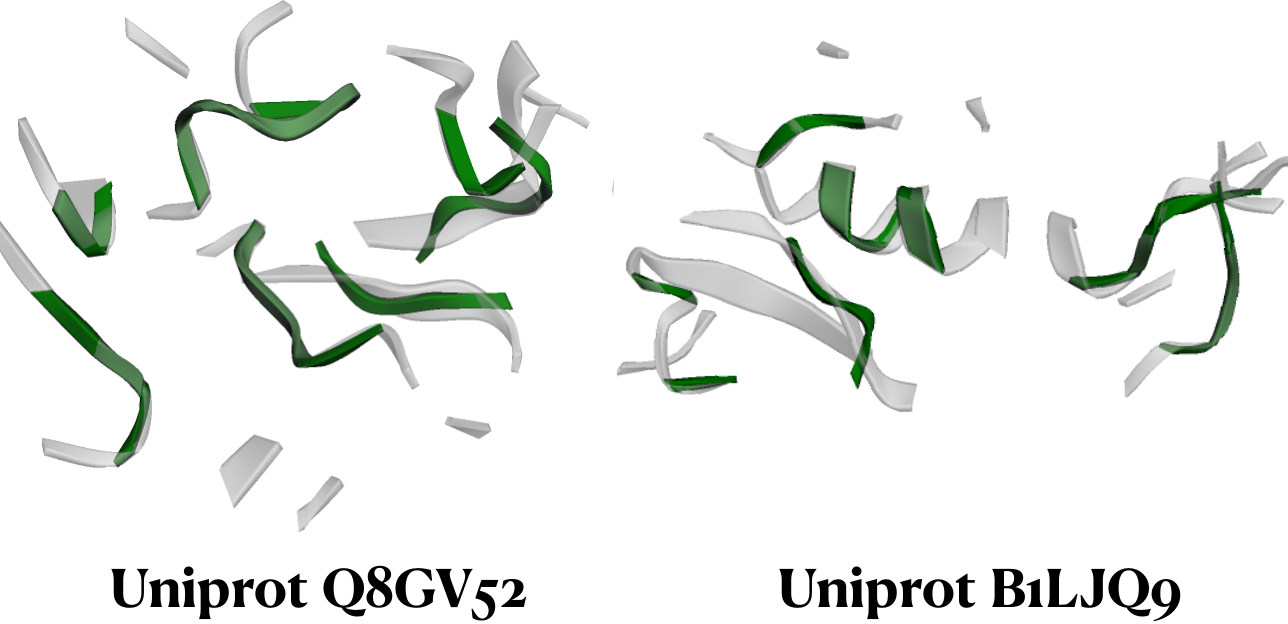}}
{
\vspace{-0.3cm}
  \caption{\textsc{GENzyme} use catalytic pockets with $64$ residues for training. \textcolor{Gray}{Grey color shows pockets with $64$ residues}, \textcolor{OliveGreen}{green color show pocket with $32$ residues}.}
  \label{fig:64residue}
  \vspace{-0.3cm}
}
\end{figure*}

In \textsc{GENzyme}, enzyme-reaction pairs with fewer than $64$ pocket residues are excluded from training. While EnzymeFlow used pockets with $32$ residues (their choice is based on LigandMPNN \citep{dauparas2023atomic}), the smaller pocket size can result in weaker training signals for learning structural information, such as pairwise distances and angles. Therefore, \textsc{GENzyme} opts for $64$-residue pockets to enhance the model's ability to capture structural details in optimization, as shown in Fig.~\ref{fig:64residue}. \textsc{GENzyme} uses a total of $128,940$ enzyme-reaction pairs with pocket structures and reaction SMILES for training.

\subsection{Pocket Generation and Pocket Sequence Design Module}
Conditioned on the input catalytic reaction \(m_r = [m_1, m_2]\), the pocket generation module of \textsc{GENzyme} generates a catalytic pocket structure \(\mathbf{E}^P\):
\small
\begin{equation}
   \textcolor{RoyalBlue}{\textsc{PocketStruct $\mathbf{E}^P$} \leftarrow \texttt{PocketGenModule}(\textsc{Substrate $m_1$, Product $m_2$})}.
\end{equation}
\normalsize
The pocket generation module leverages EnzymeFlow \citep{hua2024enzymeflow} to generate catalytic pocket structures. Originally, EnzymeFlow is a conditional flow matching model that co-generates pocket structures, amino acid types, enzyme-reaction co-MSAs, and enzyme commission numbers.

In \textsc{GENzyme}, we first generate catalytic pocket structures \(\mathbf{E}^P\), while the pocket sequences \(\mathbf{A}^P=\{a_1,\dots,a_{N_P}\}\) is co-generated using a pocket inverse folding head. However, during training, we still encode and predict masked amino acid types and enzyme-reaction co-MSAs for the purpose of masked autoencoder loss, while enzyme commission numbers are omitted. In addition to flow-matching losses on SE(3) frames and pairwise distance losses from \cite{hua2024enzymeflow}, \textsc{GENzyme} incorporates more restricted geometric regularization techniques, including FAPE loss, LDDT loss, TM loss, and violation loss as proposed in \cite{jumper2021highly, ahdritz2024openfold}, to enhance the structural accuracy of generated pockets:
\small
\begin{equation}
\begin{split}
\mathcal{L}_\text{FAPE} &= \frac{1}{N_P} \sum_{i=1}^{N_P} \left\| E^P_i \mathbf{p}_i - \hat{E}^P_i \hat{\mathbf{p}}_i \right\|^2,
\mathcal{L}_\text{LDDT} = \frac{1}{N_P} \sum_{i=1}^{N_P} \left( \frac{1}{|\text{N}(i)|} \sum_{j \in \text{N}(i)} \frac{1}{1 + \left( \frac{d_{ij} - \hat{d}_{ij}}{d_0} \right)^2} \right), 
\\
\mathcal{L}_\text{TM} &= \frac{1}{N_P} \max \left( \sum_{i=1}^{N_P} \frac{1}{1 + \left( \frac{d_i}{d_0} \right)^2} \right).
\end{split}
\end{equation}
\normalsize
Here, in \(\mathcal{L}_\text{FAPE}\), \(\hat{E}^P\) denotes the predicted pocket residue frame, and \(\mathbf{p}_i\) and \(\hat{\mathbf{p}}_i\) represent the true and predicted Euclidean atom positions in the residue frame, respectively. In \(\mathcal{L}_\text{LDDT}\), \(\text{N}(i)\) refers to the set of neighboring residues of the \(i\)-th residue, and \(d_{ij}\) and \(\hat{d}_{ij}\) are the true and predicted distances between residues \(i\) and \(j\), with \(d_0 = 10\mathring{\text{A}}\) as a threshold. In \(\mathcal{L}_\text{TM}\), \(d_i\) is the distance between the \(i\)-th residue pair in the true and predicted structures, and \(d_0\) is scaled as \(d_0 = 1.24 \sqrt{N_P - 15} - 1.8\), following \cite{jumper2021highly}. Additionally, the violation loss penalizes deviations in bond lengths, angles, and steric clashes within the predicted pocket frames.

To further improve generalizability, we relax the substrate conformation assumption from EnzymeFlow. While EnzymeFlow conditions pocket generation on pre-computed substrate conformations, we use a 2D graph representation of the substrate molecule, allowing \textsc{GENzyme} to work even when the optimal substrate conformation for a catalytic reaction is unknown. This enhances the flexibility and applicability of the model across various reactions.

\textsc{GENzyme} also generates the pocket sequence \(\mathbf{A}^P\) for the pocket structure \(\mathbf{E}^P\):
\small
\begin{equation}
\begin{split}
   \textcolor{RoyalBlue}{\textsc{PocketSeq $\mathbf{A}^P$} } & \textcolor{RoyalBlue}{\leftarrow \texttt{PocketInvFoldModule}(\textsc{PocketStruct $\mathbf{E}^P$})},
   \\
   \textcolor{RoyalBlue}{\textsc{Pocket $\mathbf{E}^P$ }} & \textcolor{RoyalBlue}{\leftarrow\texttt{Integrate}(\textsc{PocketStruct $\mathbf{E}^P$}, \textsc{PocketSeq $\mathbf{A}^P$})}.
   \end{split}
\end{equation}
\normalsize
While structure-sequence co-design has become more popular in recent protein design models, simply adding a naive amino acid prediction head to the model can be insufficient for capturing the important structural and neighborhood information necessary for accurate sequence design. An inverse folding head, on the other hand, captures more structurally meaningful information by leveraging message passing between both neighboring and long-range residues.

For sequence co-design, we enhance the model by replacing the naive amino acid prediction head with an inverse folding head powered by PiFold \citep{gao2022pifold}. PiFold excels in inference speed compared to other methods, such as ProteinMPNN \citep{dauparas2022robust} and ESM-IF \citep{lin2022language}. We pre-train the pocket inverse folding module on the CATH-4.2 dataset \citep{sillitoe2021cath} and then fine-tune it on EnzymeFill pocket data. In \textsc{GENzyme}, the fine-tuned inverse folding head is integrated for end-to-end catalytic pocket design. This involves using cross-entropy loss between the predicted sequences of the generated pockets and the true sequences of true catalytic pockets, ensuring accurate sequence generation and refinement.

\subsection{Catalytic Pocket Inpatining and Enzyme Inverse Folding Module}

Following the catalytic pocket generation, where the active site predominantly determines the enzyme function, \textsc{GENzyme} inpaints the pocket structure \(\mathbf{E}^P\) into a full enzyme structure \(\mathbf{E}\):
\small
\begin{equation}
   \textcolor{RoyalBlue}{\textsc{EnzymeStruct $\mathbf{E}$ }} \textcolor{RoyalBlue}{\leftarrow\texttt{PocketInpaintModule}(\textsc{Pocket $\mathbf{E}$})}.
\end{equation}
\normalsize
While the catalytic pocket is important to enzyme function—serving as the active site for substrate binding and catalysis—inpainting the catalytic pocket into a full enzyme structure is essential for overall enzyme stability, proper folding, and determining other properties \citep{robinson2015enzymes}. \textsc{GENzyme} fine-tunes ESM3 \citep{hayes2024simulating} on    EnzymeFill for catalytic pocket inpainting, it can inpaint catalytic pockets and predict enzyme sequences through Gibbs or discrete diffusion processes \citep{lu2024structure}. Although other models like RFDiffusion \citep{wang2022scaffolding} and Genie2 \citep{lin2024out} could also be employed\footnote{Although developing a pocket inpainting model from scratch would be novel from an algorithmic standpoint, it is not necessary, as these protein models are well-trained and optimized on millions of protein representations. Fine-tuning them for downstream applications, \textit{e.g.,} enzymes, is sufficient for domain-specific usage.}. ESM3 is chosen for its faster inference speed, and we are exploring an additional approach of training our own generative inpainting model.

\textbf{EnzymeESM with Pocket-specified Hierarchical Fine-tuning.}
To enhance performance on enzyme structures, it is necessary to fine-tune existing large protein models on catalytic pockets for inpainting, as current models may randomly mask residues during training, including those that are less functionally relevant and not involved in catalysis. Therefore, we introduce EnzymeESM, which fine-tunes ESM3 on the EnzymeFill dataset with a hierarchical fine-tuning approach. Initially, we fine-tune the model by randomly masking residues across the entire enzyme structure, followed by targeted fine-tuning where only the residues that are not belong to catalytic regions are masked for training. In \textsc{GENzyme}, this fine-tuned EnzymeESM is integrated for end-to-end catalytic pocket inpainting. During training, we not only optimize the catalytic pocket to match the true pocket but also teach-forcing the inpainted full enzyme to align with the true enzyme structure. This dual optimization enables \textsc{GENzyme} to generate full enzymes with improved stability, proper folding, and enhanced pocket specificity.

Additionally, EnzymeESM is used to co-design pocket sequences over the full enzyme structure and is further optimized in \textsc{GENzyme} for end-to-end enzyme generation, as:
\small
\begin{equation}
\begin{split}
   \textcolor{RoyalBlue}{\textsc{EnzymeSeq $\mathbf{A}$ }} & \textcolor{RoyalBlue}{\leftarrow\texttt{EnzymeInvFoldModule}(\textsc{EnzymeStuct $\mathbf{E}$})},                \\
   \textcolor{RoyalBlue}{\textsc{Enzyme $\mathbf{E}$ }} & \textcolor{RoyalBlue}{\leftarrow\texttt{Integrate}(\textsc{EnzymeStuct $\mathbf{E}$}, \textsc{EnzymeSeq $\mathbf{A}$})}.
   \end{split}
\end{equation}
\normalsize
\textsc{GENzyme} designs enzymes by generating catalytic pockets and inpainting them to full structures through dual optimization. This approach ensures that \textsc{GENzyme} accurately generates both the enzyme structure and sequence, optimizing substrate specificity and enabling the catalysis for potentially unseen reactions.

\subsection{Pocket-specific Enzyme-Substrate Binding Module}
\textsc{GENzyme} predicts and optimizes the substrate molecule conformation \( m_1 \) given the catalytic pocket \(\mathbf{E}^P\), and followed by generating the enzyme-substrate complex \(\mathbf{C}\), as:
\small
\begin{equation}
   \textcolor{RoyalBlue}{\textsc{Complex $\mathbf{C}$ } \leftarrow \texttt{BindingModule}(\textsc{Enzyme $\mathbf{E}$}, \textsc{Substrate $m_1$})}.
\end{equation}
\normalsize
The enzyme-substrate binding module is the only component in \textsc{GENzyme} that is not trained end-to-end. However, during sampling and inference, the binding module operates end-to-end. It cannot be trained this way due to current data limitations: the substrate conformations computed by AlphaFill are not guaranteed optimal, thus directly using these conformations can result in binding poses with serious atomic clashes. Addressing this data limitation—specifically, identifying the optimal enzyme-substrate binding poses—is a key challenge that \textsc{GENzyme} aims to solve.
Once the binding poses are validated as being at their optimal geometries\footnote{Although \textsc{GENzyme} can predict enzyme-substrate binding poses with computationally optimal affinities, experimental validation, such as wet-lab assays, is required to ensure the predicted binding poses are within an acceptable RMSD range (\textit{e.g.,} $2\mathring{\text{A}}$), similar to how DeepMind demonstrated the accuracy of AlphaFold-predicted structures through RMSD and distance evaluation.}, we can establish a feedback loop, using data generated by \textsc{GENzyme} to train an enzyme-substrate co-generation model like AlphaFold3 \citep{abramson2024accurate} using techniques like diffusion or flow matching.

\textsc{GENzyme} incorporates Uni-Mol Docking v2 \citep{alcaide2024uni} in its pocket-specific enzyme-substrate binding module. Uni-Mol Docking v2 is selected for its strong performance in predicting binding poses with low distance errors, consistently outperforming AlphaFold-latest on benchmarks like PoseBusters \citep{buttenschoen2024posebusters}. Within the binding module, multiple substrate conformations are computed and further optimized within the catalytic pocket. \textsc{GENzyme} then outputs the predicted enzyme-substrate complex with the lowest affinity scores.

\section{Visualization}

\subsection{Baselines}
\begin{figure*}[ht!]
\vspace{-0.2cm}
\centering
{
\includegraphics[width=1\textwidth]{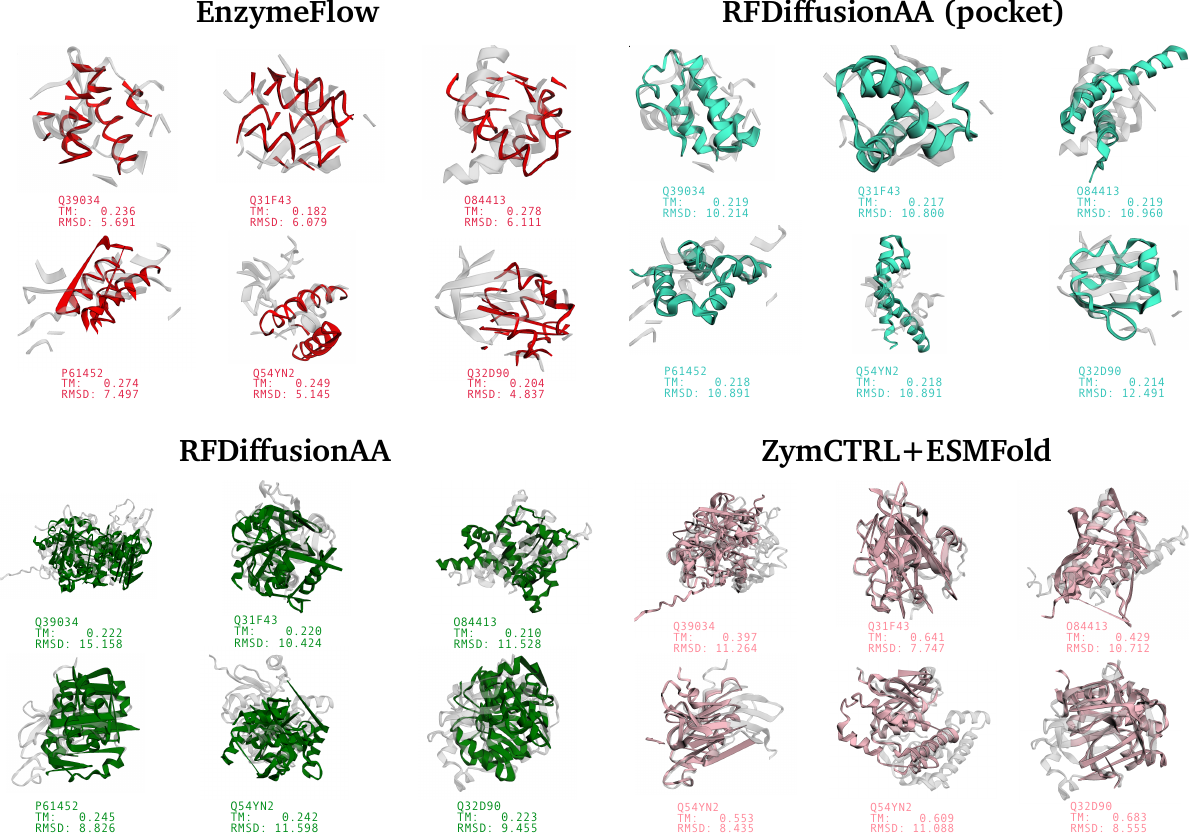}}
{
\vspace{-0.5cm}
  \caption{Baseline visualizations including EnzymeFlow, RFDiffusionAA (pocket), RFDiffusionAA, and ZymCTRL+ESMFold. These visual examples are selected by best \texttt{TM-score}. \textcolor{gray}{The ground-truth samples are shown in gray}, and generated samples are colored.}
  \label{fig:baseline.visual}
  \vspace{-0.3cm}
}
\end{figure*}

\subsection{Pocket Design with GENzyme}
\label{app:alphaenzyme.pocket.design}
\begin{figure*}[ht!]
\vspace{-0.3cm}
\centering
{
\includegraphics[width=1\textwidth]{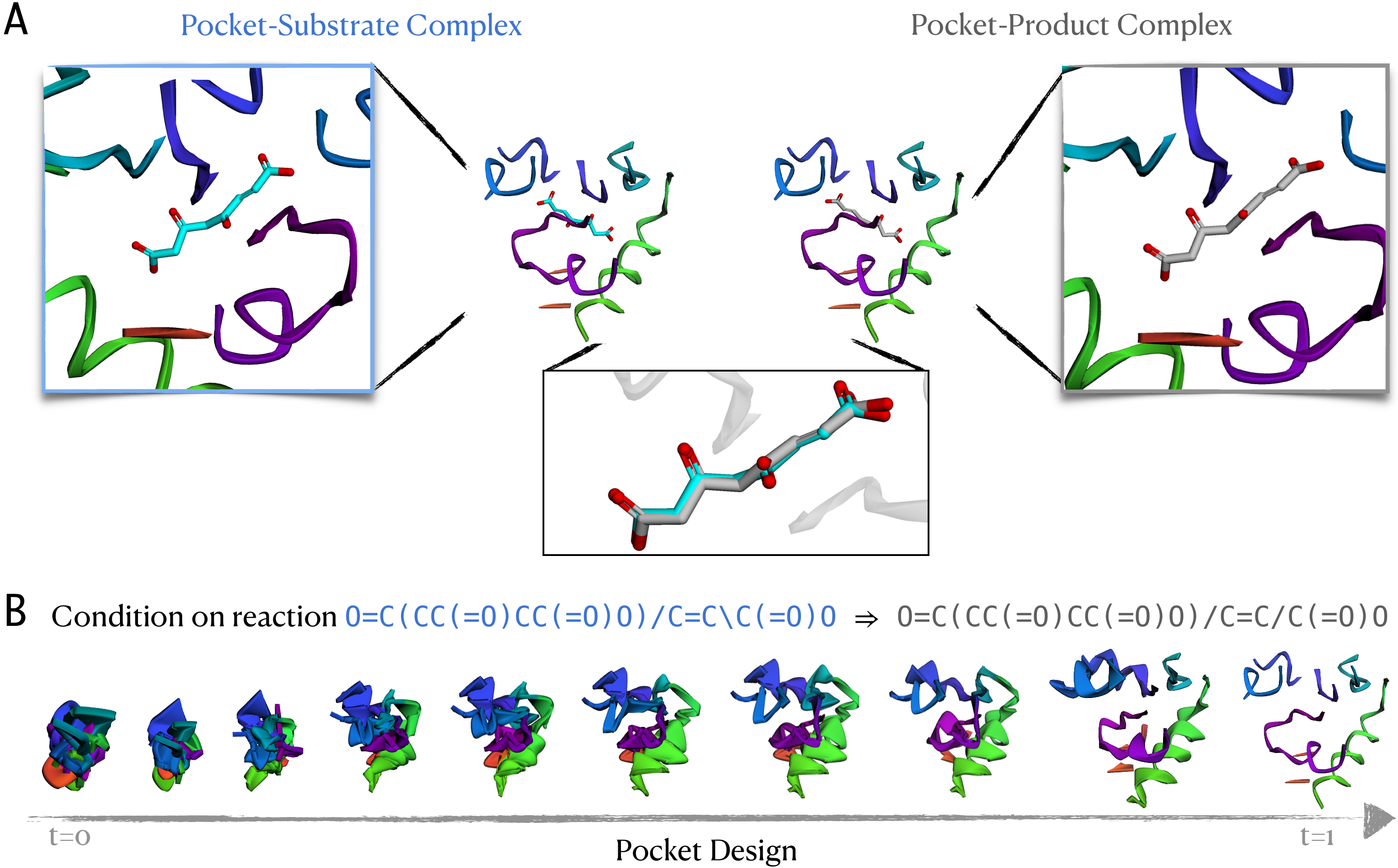}}
{
\vspace{-0.7cm}
  \caption{Catalytic Pocket Design with \textsc{GENzyme}.}
  \label{fig:pocket_example2}
  \vspace{-0.2cm}
}
\end{figure*}

\begin{figure*}[ht!]
\centering
{
\includegraphics[width=1\textwidth]{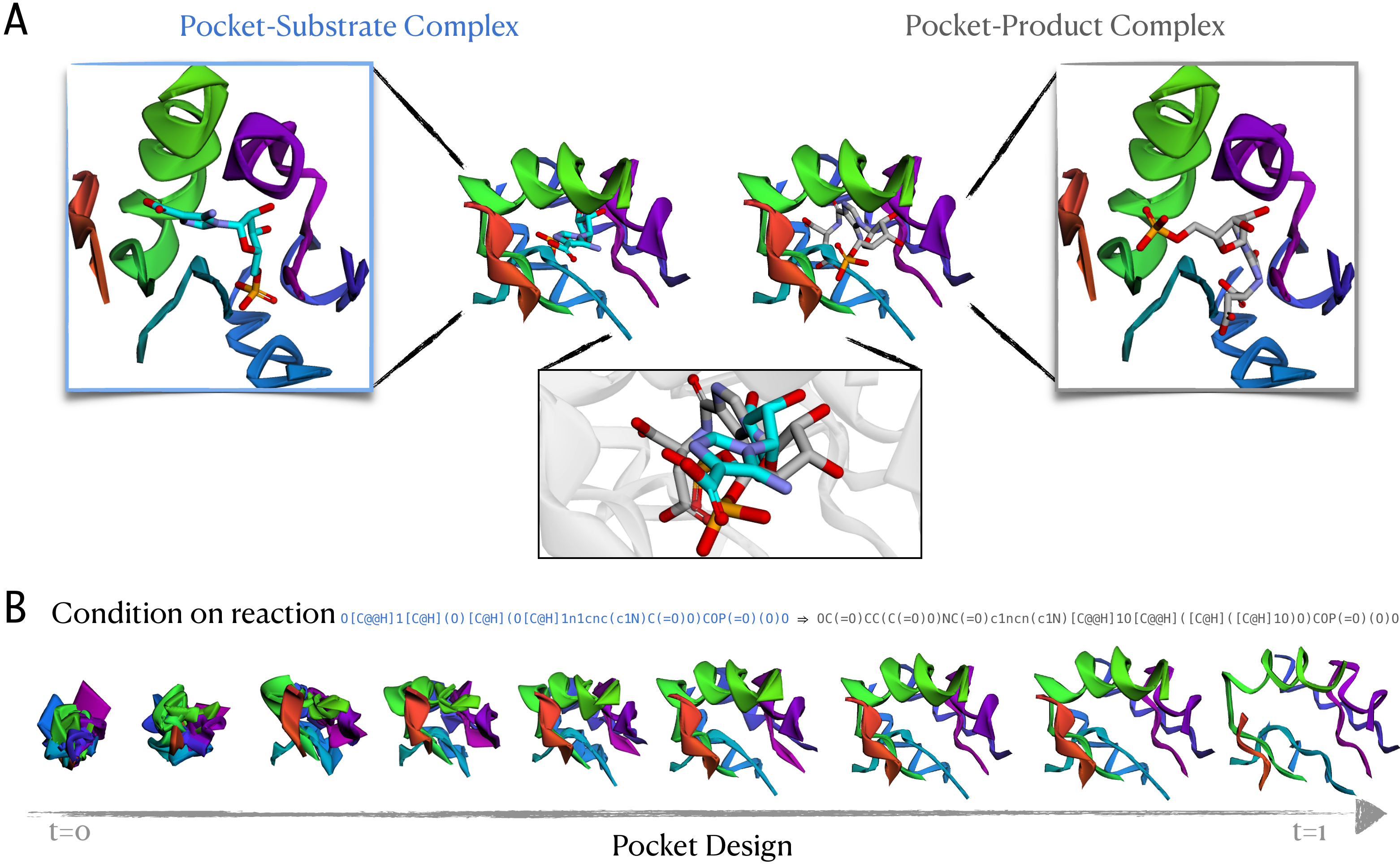}}
{
\vspace{-0.7cm}
  \caption{Catalytic Pocket Design with \textsc{GENzyme}.}
  \label{fig:pocket_example3}
}
\end{figure*}

\begin{figure*}[ht!]
\centering
{
\includegraphics[width=1\textwidth]{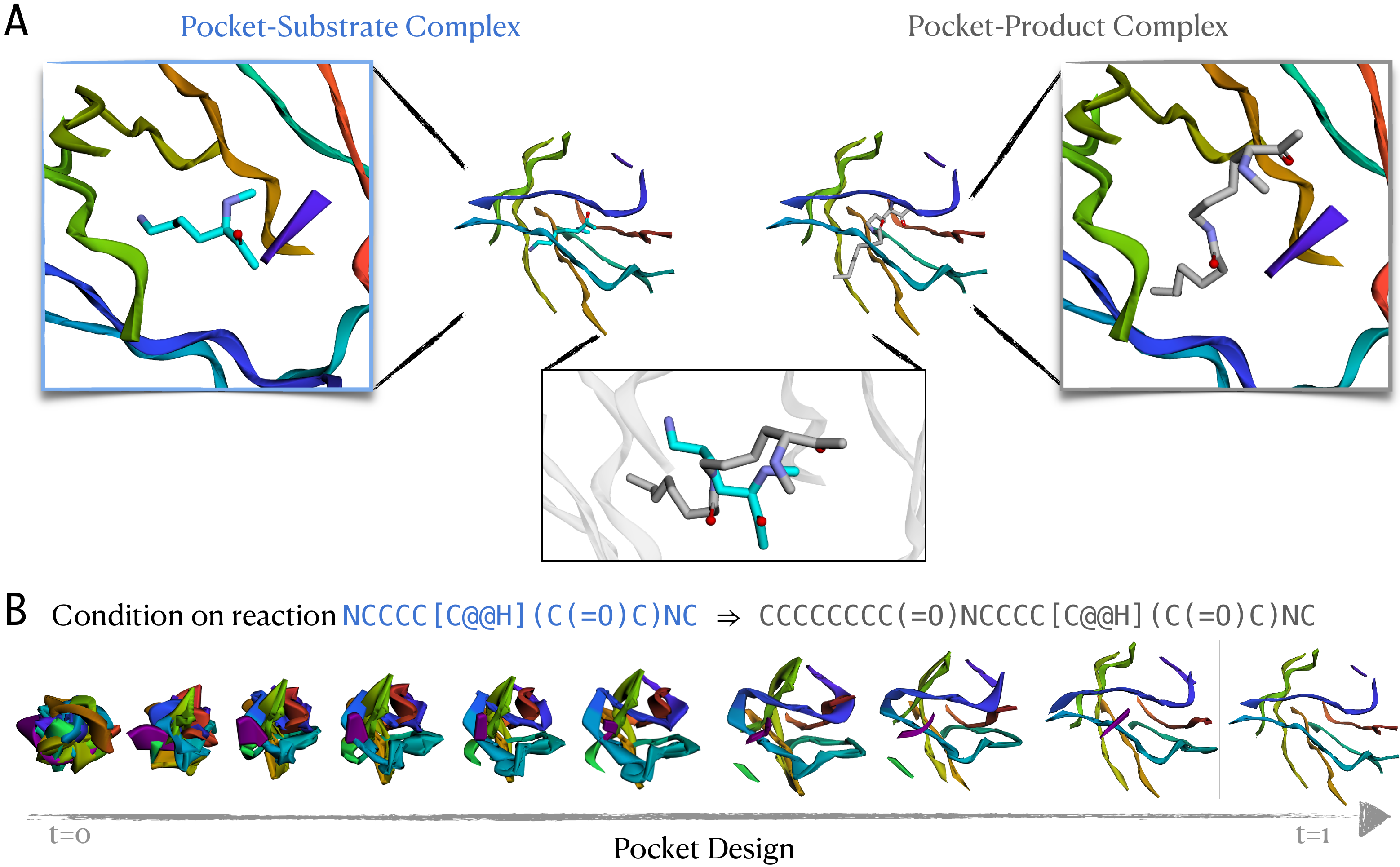}}
{
\vspace{-0.7cm}
  \caption{Catalytic Pocket Design with \textsc{GENzyme}.}
  \label{fig:pocket_example4}
  \vspace{-0.2cm}
}
\end{figure*}

\end{document}